\documentclass[12pt]{article}
\usepackage[a4paper]{geometry}
\usepackage{amsmath}
\usepackage{amsfonts}
\usepackage[english]{babel}
\usepackage{amssymb}
\usepackage{amsthm}
\usepackage{hyperref}
\usepackage{amstext}
\usepackage{caption}
\usepackage{etoolbox}
\usepackage{makeidx}
\usepackage{amsfonts}
\usepackage{sectsty}
\usepackage{amscd}
\usepackage{mathrsfs}
\usepackage{dsfont}
\usepackage{enumitem} 
\usepackage{braket}
\usepackage{authblk}
\usepackage{amsmath,empheq}
\usepackage{subfig}

\usepackage[]{latexsym}

\usepackage{soul}

\usepackage{fancyhdr}
\usepackage{physics}
\usepackage{eufrak}
\usepackage[small,nohug,heads=littlevee]{diagrams}
\usepackage{circuitikz}
\diagramstyle[labelstyle=\scriptstyle]

\newcommand{\be}{\begin{equation}}
\newcommand{\ee}{\end{equation}}
\newcommand{\bes}{\begin{equation}\begin{split}}
\newcommand{\ees}{\end{split}\end{equation}}

\newcommand{\bea}{\begin{eqnarray}}
\newcommand{\eea}{\end{eqnarray}}
\newcommand{\vsp}{\vspace{0.4cm}}

\newcommand{\stsph}{\mathcal{S}(\mathcal{H})}

\newcommand{\hh}{\mathcal{H}}
\newcommand{\bh}{\mathcal{B}(\mathcal{H})}
\newcommand{\Glh}{\mathcal{GL}(\mathcal{H})}

\newcommand{\Uh}{\mathcal{U}(\mathcal{H})}
\newcommand{\uh}{\mathfrak{u}(\mathcal{H})}

\newcommand{\SUh}{\mathcal{SU}(\mathcal{H})}

\newcommand{\suh}{\mathfrak{su}(\mathcal{H})}

\newcommand{\traceh}{\mathfrak{T}^{1}(\mathcal{H})}

\newcommand{\trr}{\textit{Tr}}
\newcommand{\lag}{\mathfrak{L}}

\newtheorem{remark}{Remark}

\newcommand{\ncd}{\newcommand}
	\ncd{\mrm}    {\mathrm}
	\ncd{\beq} {\begin{equation}}
	\ncd{\eeq} {\end{equation}}

	\def\d{{\rm d}}
	
	\def\i{{\rm i}}

	\def\E{\mathscr{E}}
	
	\def\S{{\mathcal{S}}}
	\def\LL{{\mathcal{L}}}
	\def\FF{{\mathcal{F}}}
	
	\def\F{{\mathfrak{F}}}
	\def\X{{\mathfrak{X}}}

	\newtheorem{example}{Example}

\title{Contact manifolds and dissipation, classical and quantum}

\author[1,5]{F. M. Ciaglia}
\author[2,6]{H. Cruz}
\author[3,4,7]{G. Marmo}

\affil[1]{\textit{\footnotesize Max Planck Institute for Mathematics in the Sciences, Inselstrasse 22, 04103 Leipzig, Germany}}

\affil[2]{\textit{\footnotesize Institut f\"ur Theoretische Physik, JW Goethe-Universit\"at Frankfurt am Main, Max-von-Laue-Str. 1, D-60438 Frankfurt am Main, Germany.}}

\affil[3]{\textit{\footnotesize Dipartimento di Fisica ``Ettore Pancini'', Universit\`{a} di Napoli ``Federico II'', Complesso Universitario di Monte S. Angelo, via Cintia, I-80126 Napoli, Italy}}

\affil[4]{\textit{\footnotesize INFN - Sezione di Napoli, Complesso Universitario di Monte S. Angelo, via Cintia, I-80126 Napoli, Italy}}

\affil[5]{\footnotesize personal e-mail:\texttt{florio.m.ciaglia@gmail.com}}

\affil[6]{\footnotesize \texttt{prado@itp.uni-frankfurt.de}}

\affil[7]{\footnotesize \texttt{marmo@na.infn.it}}

\date{}

\begin{document}

%
%
%

\maketitle

\begin{abstract}
Motivated by a geometric decomposition of the vector field associated with the Gorini--Kossakowski--Lindblad--Sudarshan (GKLS) equation for finite-level open quantum systems, we propose a generalization of the recently introduced contact Hamiltonian systems for the description of dissipative-like dynamical systems in the context of (non-necessarily exact) contact manifolds.
In particular, we show how this class of dynamical systems naturally emerges in the context of Lagrangian Mechanics and in the case of nonlinear evolutions on the space of pure states of a finite-level quantum system. 
\end{abstract}

\section{Introduction}

The advent of Quantum Technologies has boosted the interest in the foundational and
phenomenological aspects of quantum mechanics. 
In particular, entanglement, one of the most important characteristic features of quantum mechanics, has stressed the role of environment in our description of physical systems. 
In general, the description of the dynamical evolution of a given system, when this system is correctly thought of as a subsystem of a larger one and the evolution is obtained as a suitable reduction of the evolution of the larger system, is in terms of density states or ``density matrices'' or ``density operators'' undergoing a transformation which is no more unitary, and we say the system is dissipative.

Dissipation is playing an increasing role in the description of dynamical systems.
This is in part due to the interest in describing open quantum systems in relation with many physical applications like quantum computing,  quantum information and quantum thermodynamics.
While in classical physics the ``coupling'' with the environment may be idealized to be negligible in the consideration of fundamental aspects, in the quantum realm the situation is more complex because of the probabilistic-statistical interpretation which give rise to  indetermination relations and other inequalities.

Accepted dynamical maps are usually required to be completely positive trace preserving maps.
If additional assumptions on the coupling with the environment are made, the evolution is described by a semigroup\footnote{This particular dynamical evolutions are also referred to as being Markovian. Recently, non-Markovian dynamics have attracted a considerable attention.
We shall, for the time being restrict our attention to Markovian dynamics.} associated with the Gorini--Kossakowski--Lindblad--Sudarshan (GKLS) equation \cite{gorini_kossakowski_sudarshan-completely_positive_dynamical_semigroups_of_N-level_systems, lindblad-on_the_generators_of_quantum_dynamical_semigroups}.
This semigroup is not isospectral nor it  preserves the rank of the state.
In generic conditions, one may start with a pure state and asymptotically reach a totally mixed state where ``correlations'' among components are totally absent, in which case we say that the state undergoes ``decoherence'' and the relative phases of the ``components'' entering the state are being ``dissipated''.

In the geometrical picture of quantum mechanics, the infinitesimal generator of the GKLS semigroup is given by an affine vector field defined on a suitable carrier space associated with the space of quantum states \cite{carinena_clemente-gallardo_jover-galtier_marmo-tensorial_dynamics_on_the_space_of_quantum_states, ciaglia_dicosmo_ibort_laudato_marmo-dynamical_vector_fields_on_the_manifold_of_quantum_states, ciaglia_dicosmo_laudato_marmo-differential_calculus_on_manifolds_with_boundary.applications}.
This carrier space is equipped with two tensor fields, a skew symmetric one which gives a Poisson brackets, and a symmetric one which is related with a symmetric Jordan product and allows for an interpretation of the indetermination relations.
It is then natural to try to analyze GKLS vector fields with respect to the additional structures on the carrier space for quantum states.
What is found is that every GKLS vector field decomposes into the sum of three vector fields: one which is a Hamiltonian vector field with respect to the Poisson bivector determining the Poisson bracket, one which is a Gradient-like vector field associated with the symmetric tensor, and  a third one which is a ``Jump'' vector field responsible for the change of rank of quantum states during the dynamical evolution.
In particular, if the quantum system under investigation and the environment are ``completely decoupled'' (e.g., the system is isolated), it is found that the Gradient-like and the Jump vector fields vanish.

The decomposition of the GKLS vector field allows us to interpret the Hamiltonian part as a ``reference dynamics'' which is conservative in the sense that it preserves some Hamiltonian function, while the sum of the Gradient-like and the Jump vector fields is considered as a ``perturbation term'' associated with dissipation.
In this sense, we see that the very concept of dissipation is not associated with the GKLS vector field itself, but, rather, with the decomposition of this vector field in terms of the relevant geometrical structures under consideration.
In this contribution we want to point out that this reformulation of the notion of dissipation should also be considered when dealing with classical systems.
Specifically, based on the geometrical description of open quantum systems, we will argue that dissipation has to be understood as a relational concept (see also \cite{cantrijn-vector_fields_generating_invariants_for_classical_dissipative_systems, deritis_marmo_platania_scudellaro-inverse_problem_in_classical_mechanics_dissipative_systems}).
Essentially, we always need some reference structure, which clearly depends on the specific physical interpretation of the dynamical system under investigation, in relation to which we may define dissipation.
Dissipation, {\itshape per se}, makes no sense.

Guided by Dirac's ``principle of analogy'', we will analyze dynamical systems in the classical setting to understand if and when they are ``conservative'' or ``dissipative'' with respect to the relevant geometrical structures.
In the Newtonian approach, an ``effective'' way to describe the coupling with the environment, say the collision of the projectile  with molecules of the fluid in which it moves, is by means of phenomenological forces which would model the effect of the collisions.
As evolution in quantum mechanics of isolated systems is described by means of the Hamiltonian formalism associated with a Poisson bracket, it would be quite natural to look at  ``dissipative'' aspects in the classical setting within the description of dynamical systems on carrier spaces endowed with symplectic or related structures.
In particular, we would like to consider some recent proposals to consider ``dissipative'' systems as described by means of Hamiltonian contact vector fields \cite{bravetti_cruz_tapias-contact_hamiltonian_mechanics}.
We shall see that this approach would cover but a very limited family of ``dissipative forces'', and we will show how we can improve the formalism exposed in \cite{bravetti_cruz_tapias-contact_hamiltonian_mechanics} and make few considerations on the possibility of obtaining ``classical dissipative systems''  in analogy with quantum dissipative ones.

In seeking for this analogy, we have to distinguish the description on the space of pure states or even wave functions (which is the usual starting point for the transition to the classical setting in the framework of Hamilton--Jacoby theory) from the description on the space of mixed states which is the natural space when the evolution is obtained from the one of the composite system. 

In the former case, we will see that it is possible to consider a non-unitary dynamical evolution associated with a Hamiltonian vector field implementing Schr\"odinger equation plus a gradient vector field playing the role of a dissipative perturbation.
Similar considerations were made in \cite{rajeev-a_canonical_formulation_of_dissipative_mechanics_using_complex-valued_hamiltonians}.
In this case, the ``Jump'' vector field does not play any role because we do not want to leave the space of pure quantum states, that is, we do not want to change the rank of the quantum states we are considering.
This non-unitary dynamics is associated with a group action, specifically, to an action of the complexification of the unitary group $\Uh$ of the Hilbert space $\hh$ of the system, that is, the complex general linear group $\Glh$.
Quite interestingly, we will be able to write these dynamical evolutions as contact Hamiltonian systems (according to the definition given in Appendix A) on the  $(2n-1)$-dimensional sphere $S^{2n-1}$ of normalized vectors in $\hh$.
This shows that in quantum mechanics there is also room for contact structures if we want to implement the normalization condition of the wave function.

In the latter case, it would be more natural to work in terms of Wigner functions on phase space which would represent mixed states.
We will not have much to say on these aspects in this paper, but we shall furnish some remarks and comments in order to properly formulate and deal with these problems.
A more articulated presentation is presently being investigated and will appear elsewhere.

\section{Geometrical description of the GKLS equation and dissipation}
\label{section: Geometrical description of the GKLS equation and dissipation}

Here we want to review the geometrical description of open quantum systems in terms of a vector field $\Gamma$ associated with the generator $\mathbf{L}$ of the GKLS equation  as done in \cite{carinena_clemente-gallardo_jover-galtier_marmo-tensorial_dynamics_on_the_space_of_quantum_states, ciaglia_dicosmo_ibort_laudato_marmo-dynamical_vector_fields_on_the_manifold_of_quantum_states, ciaglia_dicosmo_laudato_marmo-differential_calculus_on_manifolds_with_boundary.applications}.
The carrier space of the dynamics is the affine space $\traceh$ of self-adjoint linear operators on $\hh$ having trace equal to $1$, where $\hh$ is the finite-dimensional, complex Hilbert space with $\mathrm{dim}(\hh)=n$ associated with the quantum system under investigation.
As it is well-known, the space $\stsph$ of quantum states of the system is the convex body in $\traceh$ composed of positive elements.
Denoting with $\{\tau_{j}\}_{j=1,...,(n^{2}-1)}$ an orthonormal set of linearly independent, traceless, self-adjoint operators in $\bh$, every element $\xi$ in $\traceh$ may be written as

\be
\xi\,=\,\frac{1}{n}\,\mathbb{I} + x^{j}\,\tau_{j}\,,
\ee
where $\mathbb{I}\in\bh$ is the identity operator and $x^{j}\,=\trr\left(\xi\,\tau^{j}\right)$ with $\{\tau^{j}\}_{j=1,...,(n^{2}-1)}$ the dual set of $\{\tau_{j}\}_{j=1,...,(n^{2}-1)}$.
Note that $\{\imath\tau^{j}\}_{j=1,...,(n^{2}-1)}$ provides an orthonormal basis of the Lie algebra $\suh$ of the special unitary group $\SUh$.
Given any self-adjoint element $\mathbf{A}=a_0 \mathbb{I} + a_j \tau^j$ in $\bh$, the linear function

\be
e_{\mathbf{A}}(\xi)\,=\,\trr(\mathbf{A}\,\xi)=a_{j}\,x^{j}
\ee
is physically interpreted as the expectation value function of the observable $\mathbf{A}$.
The special unitary group $\SUh$ determines a Poisson tensor on $\traceh$ according to the formula

\be
\Lambda\,=\,c^{jk}_{l}\,x^{l}\,\,\frac{\partial}{\partial x^{j}}\,\wedge\,\frac{\partial}{\partial x^{k}}\,,
\ee
where the $c^{jk}_{l}$'s are the structure constants of the Lie algebra $\suh$ with respect to the orthonormal basis $\{\imath\tau^{j}\}_{j=1,...,(n^{2}-1)}$.
This Poisson tensor allows us to define the Hamiltonian vector fields.
In particular, we may define the Hamiltonian vector fields $X_{\mathbf{A}}$ associated with the expectation value functions $e_{\mathbf{A}}$.
These vector fields provide a realization of $\suh$ that integrates to an action of $\SUh$ given by

\be
\xi\;\mapsto\;\;\mathbf{U}\,\xi\,\mathbf{U}^\dagger\,.
\ee
The orbits of this action in the space $\stsph$ of quantum states (i.e., the integral curves in $\stsph$ of the Hamiltonian vector fields $X_{\mathbf{A}}$ as $\mathbf{A}$ freely varies in the set of self-adjoint elements in $\bh$) defines the so-called  manifolds of isospectral quantum states (see \cite{chruscinski_ciaglia_ibort_marmo_ventriglia-stratified_manifold_of_quantum_states, ciaglia_dicosmo_ibort_laudato_marmo-dynamical_vector_fields_on_the_manifold_of_quantum_states, grabowski_kus_marmo-geometry_of_quantum_systems_density_states_and_entanglement, grabowski_kus_marmo-symmetries_group_actions_and_entanglement}).
Given a quantum state $\rho\in\stsph$, the GKLS equation describes the most general form  of the infinitesimal generator $\mathbf{L}$ for a semigroup of completely-positive, trace-preserving maps (\cite{gorini_kossakowski_sudarshan-completely_positive_dynamical_semigroups_of_N-level_systems, lindblad-on_the_generators_of_quantum_dynamical_semigroups})

\be
\mathbf{L}(\rho)\,=\,\imath\left[\rho,\,\mathbf{H}\right] -2\,\rho\,\odot\mathbf{V} + \sum_{j=1}^{R}\,\mathbf{v}_{j}\,\rho\,\mathbf{v}^{\dagger}_{j}\,,
\ee
where $R\leq (n^{2}-1)$,  $\mathbf{H}$ is a self-adjoint element in $\bh$, $\mathbf{V}=\sum_{j=1}^{R}\mathbf{v}_{j}^{\dagger}\mathbf{v}_{j}$, $[\cdot,\,\cdot]$ and $\odot$ denote, respectively, the commutator and the anticommutator product in $\bh$.
Recently (\cite{carinena_clemente-gallardo_jover-galtier_marmo-tensorial_dynamics_on_the_space_of_quantum_states, ciaglia_dicosmo_ibort_laudato_marmo-dynamical_vector_fields_on_the_manifold_of_quantum_states, ciaglia_dicosmo_laudato_marmo-differential_calculus_on_manifolds_with_boundary.applications}), it has been shown that the dynamics generated by the GKLS equation may be described in terms of an affine vector field $\Gamma$ on $\traceh$.
This vector field may be decomposed in terms of a Hamiltonian vector field $X_{\mathbf{H}}$, a Gradient vector field $Y_{\mathbf{V}}$ and a Jump vector field $Z_{\mathcal{K}}$ as follows

\be\label{eqn: GKLS vector field I}
\Gamma\,=\, X_{\mathbf{H}} - Y_{\mathbf{V}} + Z_{\mathcal{K}}\,.
\ee
The explicit form of these vector fields is

\be
\begin{split}
X_{\mathbf{H}}\,&=\, c^{jk}_{l}\,H_{k}\,x^{l}\,\frac{\partial}{\partial x^{j}} \\
Y_{\mathbf{V}}\,&=\, \left(\frac{d^{jk}_{l}}{2}\,\,V_{k}\,x^{l}+   \frac{V^{j}}{n} \right)\,\frac{\partial}{\partial x^{j}} -  e_{\mathbf{V}}\,\Delta\\
Z_{\mathcal{K}}\,& =\, \left(\trr\left(\mathcal{K}(\tau_{k})\tau^{j}\right)\,x^{k}\, + \frac{\mathcal{V}^{j}}{n}\right)\, \frac{\partial}{\partial x^{j}} - e_{\mathbf{V}}\,\Delta\,,
\end{split}
\ee
with
\be
\begin{matrix} H_{j}\,=\,\trr(\mathbf{H}\tau_{j})\,, & V^{j}\,=\,\,\trr(\mathbf{V}\tau^{j})\,=\,\delta^{jk}\,V_{k}\,,\\ & \\
f_{\mathbf{V}}(\xi)\,=\,V_{j}x^{j}\,, & \mathcal{V}^{j}=\trr\left(\sum_{k=1}^{R}\mathbf{v}_{k}\mathbf{v}_{k}^{\dagger}\,\tau^{j}\right)\,, \\ & \\
c_{l}^{jk}\,=\,\imath\trr\left([\tau^{j},\,\tau^{k}]\,\tau_{r}\right)\,, & d^{jk}_{l}\,=\,\trr\left(\{\tau^{j},\,\tau^{k}\}\,\tau_{l}\right)\,,
\end{matrix}
\ee
and

\be
\mathcal{K}(\tau_{j})=\,\sum_{k=1}^{R}\,\mathbf{v}_{k}\,\tau_{j}\,\mathbf{v}_{k}^{\dagger}\,.
\ee
Note that the Gradient and Jump vector fields are nonlinear vector fields due the presence of last term $e_{\mathbf{V}}\,\Delta$ in their expression.
However, their nonlinear terms  cancel out when taking the sum and we are left with the following affine vector field

\be\label{eqn: GKLS vector field II}
\Gamma \,=\,\mathscr{A}^{j}_{l}\,x^{l}\,\frac{\partial}{\partial x^{j}} + \mathscr{B}^{j}\,\frac{\partial}{\partial x^{j}}\,=\, \left( \mathscr{H}^{j}_{l} - \mathscr{V}^{j}_{l} + \mathscr{K}^{j}_{l}   \right)\,x^{l} \,\frac{\partial}{\partial x^{j}}+ \mathscr{B}^{j}\,\frac{\partial}{\partial x^{j}}\,,
\ee
where

\be\label{eqn: decomposition of GKLS vector field}
\begin{split}
\mathscr{H}^{j}_{l}\,=\, c^{jk}_{l}\,H_{k}\,, &\;\;\;\;
\mathscr{V}^{j}_{l}\,=\, \frac{d^{jk}_{l}\, V_{k}}{2} \\
\mathscr{K}^{j}_{l}\,=\, \trr\left(\mathcal{K}(\mathbf{e}_{l})\mathbf{e}^{j}\right) \,, &\;\;\;\; \mathscr{B}^{j}\,=\, \frac{\mathcal{V}^{j} - V^{j}}{n}\,.
\end{split}
\ee
From the geometrical point of view, we have that the flow of the Hamiltonian vector field $X_{\mathbf{H}}$ preserves the spectrum of the quantum states, the flow of the Gradient vector field $Y_{\mathbf{V}}$ does not preserve the spectrum but preserves the rank and the flow of the Jump vector field $Z_{\mathcal{K}}$ does not preserve the spectrum nor the rank of quantum states (see \cite{ciaglia_dicosmo_ibort_laudato_marmo-dynamical_vector_fields_on_the_manifold_of_quantum_states} for a detailed treatment).
Furthermore, it is immediate to realize that $X_{\mathbf{H}}$ is Hamiltonian with respect to the Poisson structure $\Lambda$ on $\traceh$, and we have

\be
X_{\mathbf{H}}\,=\,\Lambda(\mathrm{d}e_{\mathbf{H}})
\ee
with $e_{\mathbf{H}}(\xi)=h_{j}\,x^{j}$.
This means that $e_{\mathbf{H}}$ is a constant of the motion for $X_{\mathbf{H}}$ which is physically interpreted as the expectation value function of the observable $\mathbf{H}$.
Both the Gradient and the Jump vector fields do not preserve $e_{\mathbf{H}}$ in the sense that the Lie derivative of $e_{\mathbf{H}}$ with respect to $Y_{\mathbf{V}}$ and $Z_{\mathcal{K}}$ is different from $0$.
Then, we may interpret the linear vector field $X_{\mathbf{H}}$ as a ``comparison dynamics'' and the affine vector field $(Y_{\mathbf{V}} + Z_{\mathcal{K}})$ as a ``perturbation term'' which is dissipating\footnote{As long as these Lie derivatives are always negative.} the observable $\mathbf{H}$.
In the quantum case, we see that the reference dynamics is selected by means of a preferred geometrical structure on the carrier space, i.e., the Poisson tensor $\Lambda$ associated with $\suh$.


\begin{remark}
\label{remark: phase damping}
It may happen that the GKLS generator $\mathbf{L}$ presents no antisymmetric part (e.g., when $\mathbf{H}=\mathbf{0}$ and all the $\mathbf{v}_{j}$'s self-adjoint or unitary) and thus $\Gamma$ has necessarily a trivial Hamiltonian contribution given by the null vector field.
In this case, since the Hamiltonian function is trivial, we have to introduce some other ``reference structure'' with respect to which the flow of $\Gamma$ may be interpreted as dissipating something.
This is the case, for instance, of the GKLS generator $\mathbf{L}$ of the so-called phase damping.
In the q-bit case ($\mathrm{dim}(\hh)=2$), we may define the phase-damping as follows.
First of all, we select an orthonormal basis $\{|1\rangle,\,|2\rangle\}$ in $\hh$ and set

\be
\begin{split}
\tau_{1}\,&=\,\frac{1}{\sqrt{2}}\,\left(|1\rangle\langle 2| + |2\rangle\langle 1|\right)\,,\\
 \tau_{1}\,&=\,\frac{\imath}{\sqrt{2}}\,\left(|1\rangle\langle 2| - |2\rangle\langle 1|\right)\,,\\
  \tau_{3}\,&=\,\frac{1}{\sqrt{2}}\,\left(|1\rangle\langle 1| - |2\rangle\langle 2|\right)\,.
\end{split}
\ee 
Then, the phase-damping generator $\mathbf{L}$ would be

\be
\mathbf{L}(\rho)=- \gamma \left(\rho - 2\tau_{3}\,\rho\,\tau_{3}\right)\,=\,-2\gamma\left(|1\rangle\langle 1|\rho|2\rangle\langle 2 | + |2\rangle\langle 2|\rho| 1\rangle\langle 1 |\right) \,,
\ee
and the explicit form of the vector field $\Gamma$ reads

\be\label{eqn: GKLS vector field phase damping qubit}
\Gamma = Z_{\mathcal{K}} = -2\gamma\left(x^{1}\,\frac{\partial}{\partial x^{1}} + x^{2}\,\frac{\partial}{\partial x^{2}}\right)\,.
\ee
From this expression, the flow $\Phi_{\tau}$ generated by $\Gamma$ is easily obtained

\be
\Phi_{\tau}(\rho)=\frac{1}{2}\,\left(\sigma_{0} + \exp(-2\gamma\tau)\left(x^{1}\sigma_{1} + x^{2}\sigma_{2}\right) + x^{3}\sigma_{3}\right)\,.
\ee
A moment of reflection shows that this dynamical evolution decreases the phase terms  of $\rho$, that is, it ``dissipates'' the off-diagonal terms of $\rho$ with respect to the orthonormal basis $\{|1\rangle,\,|2\rangle\}$ in $\hh$.
Furthermore, if we consider the Lie algebra structure on linear functions coming from $\Lambda$

\be
\{x^{j},\,x^{k}\}\,:=\,\Lambda\left(\mathrm{d}x^{j},\,\mathrm{d}x^{k}\right)\,,
\ee
we immediately see that the dynamical evolution $\Phi_{\tau}$ is such that the asymptotic bracket

\be
\left\{x^{j},\,x^{k}\right\}_{\infty}\,=\,\lim_{\tau\rightarrow+\infty}\,\left(\Phi_{\tau}^{*}\Lambda\right)\left(\mathrm{d}x^{j},\,\mathrm{d}x^{k}\right)
\ee
is a contraction of the quantum Lie algebra (\cite{alipour_chruscinski_facchi_marmo_pascazio_rezakhani-dynamical_algebra_of_observables_in_dissipative_quantum_systems, carinena_clemente-gallardo_jover-galtier_marmo-tensorial_dynamics_on_the_space_of_quantum_states, chruscinski_facchi_marmo_pascazio-the_observables_of_a_dissipative_quantum_system}), and we may say that the dynamical evolution is ``dissipating'' the Poisson tensor $\Lambda$.
This brings in the possibility of characterizing dissipation in terms of tensor fields rather than functions.
\end{remark}

\section{Contact structures and dissipative-like systems for pure quantum states}
\label{section: Contact structures and dissipative-like systems for pure quantum states}

Pure states are extreme states in the convex body of quantum matrix states. 
If we want to restrict our attention to pure states then we can no longer refer to the full expression of the GKLS vector field because the ``Jump'' vector field takes us out of the manifold of pure quantum states.
Consequently, we may consider only  the Hamiltonian and Gradient vector fields obtaining a non-unitary dynamics which is also  ``non-linear'' in the sense that it can not be described by a linear vector field on the affine hyperplane $\traceh$ dynamics. 
We will describe these dynamics on three different spaces.

First of all, we will describe it on the complex projective space $\mathbb{CP}(\hh)$ which may be identified with the manifold of pure quantum states.
In this case, the non-unitary dynamics will be described by means of a Hamiltonian vector field implementing Schr\"odinger equation plus a Gradient vector field playing the role of a dissipative perturbation.
Furthermore, this non-unitary dynamics turns out to be  associated with an action of the complex general linear group $\Glh$ on $\mathbb{CP}(\hh)$.

Then, the same dynamical evolution is described in terms of non-linear vector fields on the Hilbert space $\hh$ of the system\footnote{More precisely, these vector fields are defined on the punctured Hilbert space $\hh_0=\hh - \{\mathbf{0}\}$, that is, the Hilbert space without the null vector.} which projects onto the Hamiltonian and Gradient vector fields on $\mathbb{CP}(\hh)$.
This description is particularly well-suited for explicit computations because of the existence of global Cartesian coordinates on $\hh$.

Finally, based on the generalization of the work \cite{bravetti_cruz_tapias-contact_hamiltonian_mechanics} presented in Appendix A, a third description is given in terms of contact Hamiltonian systems on the  $(2n-1)$-dimensional sphere $S^{2n-1}$ of normalized vectors in $\hh$.

\vsp

Let us consider a finite-level quantum system with Hilbert space $\hh$, such that $\mathrm{dim}(\hh)=n$.
Selecting an orthonormal basis $\{|\mathbf{e}_{j}\rangle\}_{j=1,...,n}$ in $\hh$ we may introduce a Cartesian coordinates system $\{x^{j},\,y^{j}\}_{j=1,...,n}$ on $\hh$, namely
	\beq \label{psi-local}
	|\psi\rangle\,=\,\sum_{j=1}^{n}\,\psi^{j}\,|\mathbf{e}_{j}\rangle\,=\,\sum_{j=1}^{n}\,\left(x^{j} + \i \,y^{j}		\right)\,|\mathbf{e}_{j}\rangle\,,
	\eeq
where $|\psi\rangle$ is an arbitrary element in $\hh$. 
The geometrical approach to quantum mechanics is based on considering the realification $\hh_{\mathbb{R}}$ of the Hilbert space $\hh$ endowed with a K\"{a}hler structure, i.e., there are a symplectic form $\omega_{\hh}$, a Riemannian metric tensor $g_{\hh}$ and a complex structure $J_{\hh}$ on $\hh$ such that $\omega_{\hh}\,\circ\,J_{\hh}\,=\,g_{\hh}$.
In coordinates we have
	\beq
	\begin{split}
	\omega_{\hh}\,&=\,\delta_{jk}\,\mathrm{d}x^{j}\,\wedge\,\mathrm{d}y^{k} \\
	g_{\hh}\,&=\, \delta_{jk}\,\left(\mathrm{d}x^{j}\,\otimes\,\mathrm{d}x^{k} + \mathrm{d}y^{j}\,\otimes\,			\mathrm{d}y^{k}\right) \\
	J_{\hh}\,&=\,\mathrm{d}y^{j}\,\otimes\,\frac{\partial}{\partial x^{j}} - \mathrm{d}x^{j}\,\otimes\,\frac{\partial}		{\partial y^{j}} \,.
	\end{split} 
	\eeq
Being $g_\hh$ and $\omega_\hh$ non-degenerate, we may also consider their inverses, i.e. the contravariant tensors $G_{\hh}\,=\,g_{\hh}^{-1}$ and $\Lambda_{\hh}\,=\,\omega_{\hh}^{-1}$ that in local coordinates have the form 
	\begin{align}
	\Lambda_{\hh}\,&=\, \delta^{jk}\,\frac{\partial}{\partial x^{j}}\,\wedge\,\frac{\partial}	
	{\partial y^{k}} \label{Poisson-H} \\
	G_{\hh}\,&=\,\delta^{jk}\,\left(\frac{\partial}{\partial x^{j}}\,\otimes\,\frac{\partial}{\partial x^{k}} 
	+ \frac{\partial}{\partial y^{j}}\,\otimes\,\frac{\partial}{\partial y^{k}}\right)\,. \label{Gradient-H}
	\end{align} 
In addition, on the Hilbert space there is a natural action of the Abelian group $\mathbb{C}_{0}$ given by
	\beq	
	|\psi\rangle\;\mapsto \lambda\,|\psi\rangle\,=\,\varrho\,\mathrm{e}^{\i \, \theta}\,|\psi\rangle\,,
	\quad
	\mbox{with}
	\quad	
	\varrho>0.
	\eeq
This action may be described infinitesimally by means of two commuting linear vector fields
	\beq\label{eqn: action on the hilbert space generating the sphere and the complex projective space}
	\Gamma\,=\,x^{j}\,\frac{\partial}{\partial y^{j}} - y^{j}\,\frac{\partial}{\partial x^{j}}\, 
	\quad
	{\rm{and}}
	\quad
	\Delta=\, x^{j}\,\frac{\partial}{\partial x^{j}} + y^{j}\,\frac{\partial}{\partial y^{j}}\, ,
	\eeq
where we see that $\Delta$ is the infinitesimal generator of dilations, while $\Gamma$ is the infinitesimal generator of the multiplication by a global phase factor. 
The action of $\mathbb{C}_{0}$ on $\hh$ allows us to introduce two smooth manifolds that are particularly relevant in quantum mechanics.

To introduce these manifolds first of all we need to introduce the Hilbert space $\hh_{0} = \hh - \{\, \mathbf{0} \,\}$, i.e., the Hilbert space $\hh$ without the null vector.
Then, on this space there are two regular distributions related with $\Delta$ and $\Gamma$.
The first one is $\mathcal{D}_{1}=\{\Delta\}$, and it is immediate to see that the quotient space\footnote{The equivalence relation $\sim_{\mathcal{D}_{1}}$ is defined as follows: first of all, we find the leaves of the regular foliation associated with $\mathcal{D}_{1}$, in this case, the leaves are open lines stemming from the origin, then, we declare two points in $\hh_{0}$ to be $\sim_{\mathcal{D}_{1}}$ equivalent if they lie on the same leaf.} $\hh_{0}/\sim_{\mathcal{D}_{1}}$ may be identified with the unit sphere in $\hh$
	\beq
	S^{2n-1}\,=\,\left\{|\psi\rangle\in\hh\; | \quad \langle\psi|\psi\rangle\,=\,1\right\}\,.
	\eeq
This reflects the following chain of well-known diffeomorphisms
	\beq
	\hh_{0}\,\cong\,\mathbb{R}^{2n}_{0}\,\cong\,\mathbb{R}^{+}\,\times\,S^{2n-1}\,.
	\eeq
In the following, the projection from $\hh_{0}$ to $S^{2n-1}$ will be denoted by $\tau$, while, an element of $S^{2n-1}$ will be denoted as $|\psi)$, where $|\psi\rangle$ is a vector in $\hh_{0}$.

Now, if we consider the distribution $\mathcal{D}_{2}\,=\,\{\Gamma,\,\Delta\}$, which is involutive, it gives rise to a foliation which is regular.
The quotient space $\hh_{0}/\sim_{\mathcal{D}_{2}}$ is known as the complex projective space $\mathbb{CP}(\hh)$ associated with $\hh$.
The projection from $\hh_{0}$ to $\mathbb{CP}(\hh)$ will be denoted by $\pi$ and the elements of $\mathbb{CP}(\hh)$, denoted by $[\psi]$ with $|\psi\rangle\in\hh_{0}$, are identified with the pure states of a quantum system with Hilbert space $\hh$.
The identification follows from the map

\be
[\psi]\:\:\mapsto\:\:\:\rho_{\psi}\,:=\,\frac{|\psi\rangle\langle\psi|}{\langle\psi|\psi\rangle}\,.
\ee

Because the vector field $\Gamma$ is tangent to $S^{2n-1}$, we may consider its restriction $\Gamma_{s}$ to $S^{2n-1}$ and build the distribution $\mathcal{D}_{s}$ associated with it.
Clearly, the quotient space $S^{2n-1}/\sim_{\mathcal{D}_{s}}$ will be precisely the complex projective space $\mathbb{CP}(\hh)$, and the canonical projection from $S^{2n-1}$ to $\mathbb{CP}(\hh)$ will be denoted by $\upsilon$, and therefore it holds that $\pi\,=\,\upsilon\,\circ\,\tau$.
Finally, we arrive at the following diagram
	\begin{diagram}
	\hh_0  			&              			&		\\
	           			&    \rdTo^{\tau}  &            \\
	  \dTo_{\pi} 		&				& S^{2n-1} \\
	      				&	\ldTo_{v}		&		 \\
	 \mathbb{CP}(\hh)	&               	   	&          
	\end{diagram}

Similarly to $\hh$, the complex projective space $\mathbb{CP}(\hh)$ of pure quantum states is a K\"{a}hler manifold, that is, there are a symplectic form $\omega$, a Riemannian metric tensor $g$ (Fubini-Study metric tensor) and a complex structure $J$ on $\mathbb{CP}(\hh)$ such that $g\,=\,\omega\,\circ J$.
Because there is no global system of coordinates on $\mathbb{CP}(\hh)$, it is better to work on the Hilbert space which is parallelizable. 
We shall work with projectable tensor fields.
We  stress that while it makes sense to deal with projectable contravariant tensor fields, it does not make sense to speak of ``projectable differential forms''.
Using the projection map $\pi$, we may look at the pullback to $\hh_{0}$ of $\omega$ and $g$

\be\label{eqn: pullback to the punctured hilbert space of the symplectic form and of the fubini-study metric tensor on the complex projective space}
\begin{split}
\omega_{0}\,&=\,\pi^{*}\omega\,=\,\Im\left(\frac{\langle\mathrm{d}\psi|\mathrm{d}\psi\rangle}{\langle\psi|\psi\rangle} -\frac{\langle\mathrm{d}\psi|\psi\rangle\langle\psi|\mathrm{d}\psi\rangle}{\langle\psi|\psi\rangle^{2}}\right)\\
g_{0}\,&=\,\pi^{*}g\,=\,\Re\left(\frac{\langle\mathrm{d}\psi|\mathrm{d}\psi\rangle}{\langle\psi|\psi\rangle} -\frac{\langle\mathrm{d}\psi|\psi\rangle\langle\psi|\mathrm{d}\psi\rangle}{\langle\psi|\psi\rangle^{2}}\right)
\end{split}\,.
\ee
By direct inspection, it is possible to see that the two-form $\omega_{0}$ on $\hh_{0}$ is such that

\be
i_{\Delta}\,\omega_{0}\,=\,
i_{\Gamma}\,\omega_{0}\,=\,0
\ee
which means that there is a two-form $\omega_{s}$ on $S^{2n-1}$ such that $\omega_{0}\,=\,\tau^{*}\omega_{s}$ and such that $\omega_{s}\,=\,\upsilon^{*}\omega$.
Note that this property is also true for the symmetric tensor $g_0$.

On the other hand, if we look at the contravariant tensors $\Lambda=\omega^{-1}$ and $G=g^{-1}$, it is immediate to check that, although $\Lambda_{\hh}$ and $G_{\hh}$ are not projectable with respect to $\pi$ nor $\tau$, we may define the following projectable tensors on $\hh_{0}$

\be\label{eqn: projectable tensor fields on H}
\begin{split}
\Lambda_{\hh}^{0}\,&=\,\langle\psi|\psi\rangle\,\Lambda_{\hh} - \Gamma\,\wedge\,\Delta\\
G_{\hh}^{0}\,&=\,\langle\psi|\psi\rangle\,G_{\hh} - \Gamma\,\otimes\,\Gamma - \Delta\,\otimes\,\Delta\,.
\end{split}
\ee
These tensors are $\pi$-related with $\Lambda=\omega^{-1}$ and $G=g^{-1}$, respectively.
Also the tensors $\langle\psi|\psi\rangle\,\Lambda_{\hh}$ and $\langle\psi|\psi\rangle\,G_{\hh}$ are projectable.
As a matter of fact, both $\Lambda_{\hh}^{0}$ and $\langle\psi|\psi\rangle\,\Lambda_{\hh}$ project onto the same one on the complex projective space, and similarly for $G_{\hh}^{0}$ and $\langle\psi|\psi\rangle\,G_{\hh}$.

We notice that the conformal factor $\langle \psi | \psi \rangle$, which is necessary for the projectability of the two tensor fields, in the case of the skew-symmetric tensor, spoils the Jacobi identity. Indeed, the new bracket is not a Poisson Bracket but is rather a Jacobi bracket.
We define Jacobi bracket in Appendix A. 
At the level of $2$-forms, while the Poisson bracket is associated with a symplectic form, the Jacobi bracket may be associated with a contact structure, in our case it will be an exact contact structure (see Appendix A for a discussion on contact manifolds and their connection with Jacobi brackets).
Thus, the restriction of Schr\"odinger-type equations to the sphere of normalized vectors, $S^{2n-1}$, will be described in terms of contact manifolds instead of symplectic manifolds.
On this manifold we should consider the Jacobi brackets instead of the Poisson brackets.

\begin{remark}[Brackets]\label{remark: quantum brackets}
By means of $\Lambda_{\hh},\,G_{\hh},\,\Lambda$ and $G$ we may define the following brackets

\be
\begin{matrix}
\left\{F,\,H\right\}_{\hh}\,:=\,\Lambda_{\hh}\left(\mathrm{d}F,\,\mathrm{d}H\right)\,,& \left(F,\,H\right)_{\hh}\,:=\,G_{\hh}\left(\mathrm{d}F,\,\mathrm{d}H\right) \\ & \\
\left\{f,\,h\right\}\,:=\,\Lambda\left(\mathrm{d}f,\,\mathrm{d}h\right)\,,& \left(f,\,h\right)\,:=\,G\left(\mathrm{d}f,\,\mathrm{d}h\right)\,,
\end{matrix}
\ee
where $F,H$ are smooth functions on $\hh$, and $f,h$ are smooth functions on $\mathbb{CP}(\hh)$.
In particular, when $F=f_{\mathbf{a}},\,H=f_{\mathbf{b}}$ and $f=e_{\mathbf{a}},\,h=e_{\mathbf{b}}$ we obtain

\be
\begin{matrix}
\left\{f_{\mathbf{a}},\,f_{\mathbf{b}}\right\}_{\hh}\,=\,f_{[\mathbf{a},\,\mathbf{b}]}\,,& \left(f_{\mathbf{a}},\,f_{\mathbf{b}}\right)_{\hh}\,=\,f_{-\imath\mathbf{a}\odot\mathbf{b}} \\ & \\
\left\{e_{\mathbf{a}},\,e_{\mathbf{b}}\right\}\,=\,e_{[\mathbf{a},\,\mathbf{b}]}\,,& \left(e_{\mathbf{a}},\,e_{\mathbf{b}}\right)\,=\,e_{-\imath\mathbf{a}\odot\mathbf{b}} - e_{\mathbf{a}}\,e_{\mathbf{b}}\,,
\end{matrix}
\ee
where $[\cdot,\,\cdot]$ denotes the Lie bracket in $\uh$, $\odot$ denotes the (matrix) anticommutator, and $e_{\mathbf{a}}\,e_{\mathbf{b}}$ the pointwise product among functions.

\end{remark}

The dynamical evolution of a closed system is associated with the action of the unitary group $\Uh$ on $\mathbb{CP}(\hh)$.
This action ``comes'' from the following linear action of $\Uh$ on $\hh$

\be
|\psi\rangle\;\mapsto\;\;|\psi_{\mathbf{U}}\rangle\,:=\,\mathbf{U}\,|\psi\rangle\,,
\ee
with $\mathbf{U}\in\Uh$, that is,  the action of $\Uh$ on the space $\mathbb{CP}(\hh)$ of pure states reads

\be
[\psi]\;\mapsto\;\;[\psi_{\mathbf{U}}]\,.
\ee
Then, the Schr\"odinger equation is the infinitesimal version of the evolution equation

\be
[\psi_{t}]\,=\,[\psi_{\mathbf{U}_{t}}]
\ee
where $\mathbf{U}_{t}\,=\,\mathrm{e}^{\imath\mathbf{H}t}$ is the one-parameter group of unitary operators generated by the Hermitean Hamiltonian operator $\mathbf{H}$.
Given $\mathbf{a}$ in the Lie algebra $\uh$ of $\Uh$, we may introduce the smooth real-valued functions

\be
\begin{split}
f_{\mathbf{a}}(|\psi\rangle)\,&=\,-\imath\,\langle\psi|\mathbf{a}|\psi\rangle\\
e_{\mathbf{a}}([\psi])\,&=\,-\imath\,\frac{\langle\psi|\mathbf{a}|\psi\rangle}{\langle\psi|\psi\rangle}\,,
\end{split}
\ee
and it is possible to prove that the curves $\mathbf{U}_{t}\,|\psi\rangle$ and $[\psi_{\mathbf{U}_{t}}]$ are integral curves of the Hamiltonian vector fields

\be\label{eqn: infinitesimal generators of the actions of the unitary group}
\begin{split}
\mathbb{X}_{\mathbf{a}}\,&=\,\Lambda_{\hh}\left(\mathrm{d}f_{\mathbf{a}}\right) \\
X_{\mathbf{a}}\,&=\,\Lambda\left(\mathrm{d}e_{\mathbf{a}}\right) \, ,
\end{split}
\ee
respectively \cite{ashtekar_schilling-geometrical_formulation_of_quantum_mechanics, carinena_ibort_marmo_morandi-geometry_from_dynamics_classical_and_quantum, cirelli_mania_pizzocchero-quantum_mechanics_as_an_infinite_dimensional_Hamiltonian_system_with_uncertainty_structure, cirelli_pizzocchero-on_the_integrability_of_quantum_mechanics_as_an_infinite_dimensional_system, ercolessi_marmo_morandi-from_the_equations_of_motion_to_the_canonical_commutation_relations, kibble-geometrization_of_quantum_mechanics}, and $\mathbb{X}_{\mathbf{a}}$ and $X_{\mathbf{a}}$ are $\pi$-related.
However, we may also use $\Lambda_{\hh}^{0}$ in order to define the ``Hamiltonian'' vector field

\be
\mathbb{X}_{\mathbf{a}}^{0}\,:=\,\Lambda_{\hh}^{0}\left(\mathrm{d}\pi^{*}e_{\mathbf{a}}\right)\,=\,\mathbb{X}_{\mathbf{a}} - \pi^{*}e_{\mathbf{a}}\,\Gamma
\ee
which is again $\pi$-related with $\mathbb{X}_{\mathbf{a}}$.

Now, we observe that on $\hh$ there is also a linear action of the complex general linear group $\Glh$ on $\hh$ given by

\be\label{eqn: action of the general linear group on the hilbert space}
|\psi\rangle\;\mapsto\;\;\Phi_{\mathbf{G}}\left(|\psi\rangle\right)\,\equiv\,|\psi_{\mathbf{G}}\rangle\,:=\,\mathbf{G}\,|\psi\rangle\,,
\ee
where $\mathbf{G}\in \Glh$.
Associated with this action, we have an action $\phi$ of $\Glh$ on the space $\mathbb{CP}(\hh)$ of pure states

\be\label{eqn: action of the general linear group on the odd sphere and complex projective space}
[\psi]\;\mapsto\;\;\phi_{\mathbf{G}}\left([\psi]\right)\,:=\,[\psi_{\mathbf{G}}]\,.
\ee
These actions are such that the following diagram commutes

\be
\begin{matrix} \hh_{0} & \stackrel{\Phi_{\mathbf{G}}}{\longrightarrow} & \hh_{0} \\ \Big\downarrow\pi &  & \Big\downarrow\pi \\  \mathbb{CP}(\hh) & \stackrel{\phi_{\mathbf{G}}}{\longrightarrow} & \mathbb{CP}(\hh) \end{matrix}\,.
\ee
When $\mathbf{G}\equiv\mathbf{U}$ is actually an element of the unitary group $\Uh\subset\Glh$, we obtain the actions of $\Uh$ on $\hh$ and $\mathbb{CP}(\hh)$, respectively.

Regarding the action $\Phi$ of $\Glh$ on $\hh$, it is possible to show that the linear vector fields

\be
\mathbb{X}_{\mathbf{a}} + \mathbb{Y}_{\mathbf{b}}\,=\,\Lambda_{\hh}(\mathrm{d}f_{\mathbf{a}}) + G_{\hh}(\mathrm{d}f_{\mathbf{b}})\,,
\ee
where $(\mathbf{a} - \imath\mathbf{b})\in\mathfrak{gl}(\hh)$, provide an anti-representation of the Lie algebra $\mathfrak{gl}(\hh)$ of $\Glh$ which integrates to the action $\Phi$.
Furthermore, according to recent works \cite{chruscinski_ciaglia_ibort_marmo_ventriglia-stratified_manifold_of_quantum_states, ciaglia_ibort_marmo-differential_geometry_of_quantum_states_observables_and_evolution}, it is possible to describe the full action $\phi$ of $\Glh$ on $\mathbb{CP}(\hh)$ using the Hamiltonian vector fields $X_{\mathbf{a}}$ together with the so-called gradient vector fields $Y_{\mathbf{b}}:=J(X_{\mathbf{b}})$.
Specifically, it has been proved that the  vector fields  $X_{\mathbf{a}} + Y_{\mathbf{b}}$ provide a Lie-algebra anti-representation of $\mathfrak{gl}(\hh)$ on $\mathbb{CP}(\hh)$ which integrates to the action $\phi$.
Furthermore, it turns out that the Hamiltonian and gradient vector fields on $\mathbb{CP}(\hh)$ coincide with the restriction to the space of pure quantum states (seen as a subset of the affine hyperplane $\traceh$ of Hermitean operators with unit trace) of the Hamiltonian and Gradient vector fields on $\traceh$ (see section \ref{section: Geometrical description of the GKLS equation and dissipation}).

On the other hand, introducing the following vector fields on $\hh_{0}$

\be
\mathbb{Y}^{0}_{\mathbf{b}}\,:=\,G_{\hh}(\mathrm{d}\pi^{*}e_{\mathbf{b}})\,=\, \mathbb{Y}_{\mathbf{b}} - \pi^{*}e_{\mathbf{b}}\,\Delta\,,
\ee
it is easy to see that $\mathbb{Y}^{0}_{\mathbf{b}}$ is tangent to $S^{2n-1}$ and that $[\Gamma,\,\mathcal{Y}^{0}_{\mathbf{b}}]=0$, from which we conclude that $\mathbb{Y}^{0}_{\mathbf{b}}$ is projectable with respect to $\pi$.
In particular, $\mathbb{Y}^{0}_{\mathbf{b}}$ is $\pi$-related with the gradient vector field $Y_{\mathbf{b}}:=J(X_{\mathbf{b}})$.
It is interesting to note that the infinitesimal generator $Z_{\mathbf{a},\mathbf{b}}\,=\,X_{\mathbf{a}} + Y_{\mathbf{b}}$ of the action $\phi$ of $\Glh$ on $\mathbb{CP}(\hh)$ may be written as the unique vector field such that

\be
i_{Z_{\mathbf{a},\mathbf{b}}}\,\omega\,=\,\mathrm{d}e_{\mathbf{a}} + \alpha_{\mathbf{b}}\,,
\ee
where $\alpha_{\mathbf{b}}=J\circ\mathrm{d}e_{\mathbf{b}}$.
This way of writing $Z_{\mathbf{a},\mathbf{b}}$ allows us to think of it as being the sum of a ``reference dynamics'' which is conservative, the vector field $X_{\mathbf{a}}$ associated with $e_{\mathbf{a}}$, with a ``perturbation term'' generating dissipation, the vector field $Y_{\mathbf{b}}$ associated with $J\mathrm{d}e_{\mathbf{b}}$.
Clearly, these decomposition depends on the existence of the symplectic form $\omega$ on $\mathbb{CP}(\hh)$.

\begin{example}[The q-bit case]
To better visualize the situation, we shall consider the example of the q-bit in which $\hh \cong \mathbb{C}^2$. 
In this case, we have that the complex projective space $\mathbb{CP}(\hh) $ is diffeomorphic to a $2$-dimensional sphere embedded into a $3$-dimensional affine space.
Indeed, we recall that every quantum state $\rho$ is a density operator on $\hh$, which means that it is a non-negative Hermitean operator with trace equal to $1$.
In particular, in the q-bit case, we can write
	\beq
	\rho = \frac{1}{2}(\sigma_0 + x^j \sigma_j)\, ,
	\eeq 
where $\sigma_0\equiv\mathbb{I}$ is the identity operator on $\hh$, $\{\sigma_1, \sigma_2, \sigma_3\}$ are Pauli operators, and the $x^j$'s are Cartesian coordinates on the space of Hermitean operators (which is the linear span of $\{\sigma_0, \sigma_1, \sigma_2, \sigma_3\}$).
The non-negativity of $\rho$ is ensured by imposing the constraint $(x^1)^2 + (x^2)^2 + (x^3)^2 \geq 1$.
Furthermore, pure states are singled out by imposing the additional constraint $\rho^2 = \rho$ which is equivalent to $(x^1)^2 + (x^2)^2 + (x^3)^2 = 1$.
From this, it is clear that the space of quantum states for a two-level system is a three-dimensional solid ball in the affine hyperplane of Hermitean operators on $\hh$ with unit trace.

Given a Hermitean operator $\mathbf{A}=a^0 \sigma_0 + a^j \sigma_j$, we can write the Hamiltonian and Gradient vector field associated with $\mathbf{A}$ as
	\beq
	X_A  = \varepsilon_{jkl} \, x^j \, a^k \, \frac{\partial }{\partial x_l}
	\quad
	\mbox{and}
	\quad
	Y_A = \delta^{jk} a_k \frac{\partial }{\partial x_j} - (a_k \, x^k) \Delta \, ,
	\eeq 
where the $\varepsilon^{jk}_l$ are the structure constants of $\SUh$ and we raise and lower indices with the Euclidean metric in $\mathbb{R}^3$.
Note that, according to the work in \cite{chruscinski_ciaglia_ibort_marmo_ventriglia-stratified_manifold_of_quantum_states, ciaglia_dicosmo_ibort_laudato_marmo-dynamical_vector_fields_on_the_manifold_of_quantum_states}, the Hamiltonian vector field may be written as $X_A = \Lambda( \d f_A, \cdot )$ where $f_A = a_j u^j$ and $\Lambda$ is the Poisson tensor $\Lambda = \varepsilon_{jkl} \, x^j \frac{\partial }{\partial x_k} \wedge \frac{\partial }{\partial x_l}$ on $\traceh$, while the Gradient vector field may be written as $Y_A = G( \d f_A, \cdot )$ where $G$ is the symmetric tensor $G = \delta^{jk} \frac{\partial }{\partial x_j} \otimes \frac{\partial }{\partial x_k} - \Delta \otimes \Delta $ and $\Delta = x^j \frac{\partial }{\partial x_j}$ is the dilation vector field on $\traceh$.

The Gradient vector field $Y_A$ describes a nonlinear dynamics which is tangent to the sphere $(x^1)^2 + (x^2)^2 + (x^3)^2 = 1$ of pure quantum states, indeed
	\beq
	\pounds_{Y_A}[(x^1)^2 + (x^2)^2 + (x^3)^2] = 2 f_A [1- (x^1)^2 + (x^2)^2 + (x^3)^2]\,,
	\eeq
and this quantity vanishes on the unit sphere. 
In the q-bit case, it is easy to visualize the situation. 
Hamiltonian vector fields are tangent to ``parallels'', while gradient vector fields are tangent to ``meridians'', orthogonal to parallels, and both of them vanish at critical points of $f_A$. 
The two critical points represent ``unstable'' and ``stable'' equilibria for the gradient dynamics.

\end{example}

Now, we will prove that it is possible to describe the ``dynamical evolution'' generated by $Z_{\mathbf{a},\mathbf{b}}\,=\,X_{\mathbf{a}} + Y_{\mathbf{b}}$ in terms of the geometry of contact Hamiltonian vector fields on contact manifolds developed in Appendix A.
In particular, the contact manifold we are referring to is the $(2n-1)$-dimensional sphere $S^{2n-1}\subset\hh$ of normalized vectors in $\hh$, and the dynamical system is a generalized contact Hamiltonian system described by equations \eqref{Cont-vect-field 3} and \eqref{Cont-vect-field 4}.

Consider the following one-form $\eta_{0}$ on $\hh_{0}$ given by

\be
\eta_{0}\,:=\,\frac{J_{\hh}\left(\mathrm{d}f_{\mathbb{I}}\right)}{2r^{2}} \,=\,\Im\left(\frac{\langle\psi|\mathrm{d}\psi\rangle - \langle\mathrm{d}\psi|\psi\rangle}{\langle\psi|\psi\rangle}\right)\,=\, \frac{1}{r^{2}}\,\delta_{jk}\,\left(x^{j}\mathrm{d}y^{k} - y^{j}\mathrm{d}x^{k}\right)\,.
\ee
It is immediate to check that

\be
i_{\Delta}\,\eta_{0}\,=\,i_{\Delta}\,\mathrm{d}\eta_{0}\,=\,0\,,
\ee
which means that there is a one-form $\eta_{s}$ on $S^{2n-1}$ such that $\eta_{0}=\tau^{*}\eta_{s}$.
Clearly, $\eta_{0}$ is invariant under the action of the unitary group $\Uh$ on $\hh_{0}$.
A direct computation shows that

\be
\Theta\,=\, \eta_{0} \,\wedge\,(\omega_{0})^{n-1}\,\wedge\,\mathrm{d}r
\ee
is a volume form on $\hh_{0}$.
Since $i_{X}\,\mathrm{d}r=0$ for every $X$ which is tangent to $S^{2n-1}$, we conclude that

\be
\Omega_{0}:= \eta_{0} \,\wedge\,(\omega_{0})^{n-1}
\ee
is the pullback to $\hh_{0}$ of a volume form $\Omega_{s}$ on $S^{2n-1}$  by means of $\tau$.
According to the general theory exposed in Appendix A, we have that $\eta_{s}$ and $\omega_{s}$ give rise to a contact structure on $S^{2n-1}$.
Note that $\Gamma_{s}$ is the Reeb vector field of the contact structure $(\eta_{s},\,\omega_{s})$, that is $i_{\Gamma_{s}} \eta_{s} =1$ and $i_{\Gamma_{s}}\omega_{s}=0$.

Given a smooth function $\mathscr{F}$ and a one-form $\alpha$ on $S^{2n-1}$, we say that the vector field $\mathcal{Z}$ is the generalized contact Hamiltonian vector field associated with $\mathscr{F}$ and $\alpha$ by means of $(\eta_{s},\,\omega_{s})$ if it satisfies the following two conditions simultaneously (see equations \eqref{Cont-vect-field 3} and \eqref{Cont-vect-field 4})

\be
i_{\mathcal{Z}}\,\eta_{s}\,=\,\mathscr{F}\,,\:\:\:\mbox{ and }\:\:\: i_{\mathcal{Z}}\,\omega_{s}\,=\,\left(\mathrm{d}\mathscr{F} - \alpha\right) - \left[\left(\mathrm{d}\mathscr{F} - \alpha\right)(\Gamma_{s})\,\right]\,\eta_{s}\,.
\ee
It is possible to look at contact Hamiltonian vector fields directly on $\hh_{0}$ where we have a well-defined global differential calculus.
Indeed, we set $\widetilde{\mathscr{F}}=\tau^{*}\mathscr{F}$ and $\widetilde{\alpha}=\tau^{*}\alpha$, and look for the vector field $\mathbb{Z}$ such that

\be
i_{\mathbb{Z}}\,\mathrm{d}r\,=\,0 \,,\:\:\:\mbox{ and }\:\:\: 
i_{\mathbb{Z}}\,\eta_{0}\,=\,\widetilde{\mathscr{F}} \,,\:\:\:\mbox{ and }\:\:\: 
i_{\mathbb{Z}}\,\omega_{0}\,=\,\left(\mathrm{d}\widetilde{\mathscr{F}} - \widetilde{\alpha}\right) - \left[\left(\mathrm{d}\widetilde{\mathscr{F}} - \widetilde{\alpha}\right)(\Gamma )\,\right]\,\eta_{0}\,.
\ee
These equations assure us that $\mathbb{Z}$ is tangent to $S^{2n-1}$, and that its restriction to $S^{2n-1}$ coincides precisely with the contact Hamiltonian vector field $\mathcal{Z}$ on $S^{2n-1}$ associated with $\mathscr{F}$  and $\alpha$ by means of the contact structure $(\eta_{s} \,,\omega_{s})$.
Now, let us fix

\be
\begin{split}
\mathscr{F}_{\mathbf{a}}\,&=\,\upsilon^{*}e_{\mathbf{a}} \,,\:\:\:\mbox{ and }\:\:\:  \widetilde{\mathscr{F}}_{\mathbf{a}}=\tau^{*}\mathscr{F}_{\mathbf{a}}\,=\,\frac{f_{\mathbf{a}}}{r^{2}}\,,\\
\alpha_{\mathbf{b}}\,&=\,\upsilon^{*}J(\mathrm{d}e_{\mathbf{b}})\,,\:\:\:\mbox{ and }\:\:\: \widetilde{\alpha}_{\mathbf{b}}=\tau^{*}\alpha_{\mathbf{b}}\,,
\end{split}
\ee
in which case it follows that $\left(\mathrm{d}\widetilde{\mathscr{F}}_{\mathbf{a}} - \widetilde{\alpha}_{\mathbf{b}}\right)(\Gamma)=0$.
In this case, we will show that the vector field  $\mathbb{Z}=\mathbb{X}_{\mathbf{a}} + \mathbb{Y}^{0}_{\mathbf{b}}$ is such that

\be\label{eqn: equations defining the unitary+gradient vector field on the punctured hilbert space}
\begin{split}
i_{\mathbb{Z}}\,\mathrm{d}r\,=\,0 \,,\:\:\:\mbox{ and }\:\:\: 
i_{\mathbb{Z}}\,\eta_{0}\,=\,\widetilde{\mathscr{F}}_{\mathbf{a}} \,,\:\:\:\mbox{ and }\:\:\: i_{\mathbb{Z}}\,\omega_{0}\,=\, \mathrm{d}\widetilde{\mathscr{F}}_{\mathbf{a}} - \widetilde{\alpha}_{\mathbf{b}}\,.
\end{split}
\ee
This means that there is a vector field $\mathcal{Z}$ on $S^{2n-1}$ which is the contact Hamiltonian vector field associated with $\mathscr{F}_{\mathbf{a}}$  and $\alpha_{\mathbf{b}}$ by means of the contact structure $(\eta_{s} \,,\omega_{s})$.

The first equation in \eqref{eqn: equations defining the unitary+gradient vector field on the punctured hilbert space} is clearly satisfied because we already saw that $\Gamma,\,\mathbb{X}_{\mathbf{a}}$ and $\mathbb{Y}^{0}_{\mathbf{b}}$ are tangent to $S^{2n-1}$.
Then, we have that

\be
i_{\mathbb{Z}}\,\eta_{0}\,=\,i_{\mathbb{X}_{\mathbf{a}}}\,\eta_{0} + i_{\mathcal{Y}^{0}_{\mathbf{b}}}\,\eta_{0}\,,
\ee

\be
\begin{split}
i_{\mathbb{X}_{\mathbf{a}}}\eta_{0}\,&=\frac{\left(J_{\hh}(\mathrm{d}f_{\mathbb{I}})\right)(\mathbb{X}_{\mathbf{a}})}{2r^{2}}\,=\,\frac{\Lambda_{\hh}\left(\mathrm{d}f_{\mathbf{a}},\,J_{\hh}\mathrm{d}f_{\mathbb{I}}\right)}{2r^{2}}\,=\\
&\,=\,\frac{G_{\hh}\left(\mathrm{d}f_{\mathbf{a}},\,\mathrm{d}f_{\mathbb{I}}\right)}{2r^{2}}\,=\,\frac{f_{\mathbf{a}}}{r^{2}}\,=\,\widetilde{\mathscr{F}}_{\mathbf{a}}\,,
\end{split}
\ee

{\footnotesize
\be
\begin{split}
i_{\mathbb{Y}^{0}_{\mathbf{b}}}\eta_{0}\,&=\frac{\left(J_{\hh}(\mathrm{d}f_{\mathbb{I}})\right)(\mathbb{Y}^{0}_{\mathbf{b}})}{2r^{2}}\,=\,\frac{G_{\hh}\left(\mathrm{d}f_{\mathbf{b}} -\pi^{*}e_{\mathbf{b}}\,\mathrm{d}f_{\mathbb{I}},\,J_{\hh}\mathrm{d}f_{\mathbb{I}}\right)}{2r^{2}}\,=\\
&\,=\,\frac{\Lambda_{\hh}\left( \mathrm{d}f_{\mathbf{b}}  - \pi^{*}e_{\mathbf{b}}\,\mathrm{d}f_{\mathbb{I}},\,\mathrm{d}f_{\mathbb{I}}\right)}{2r^{2}}\,=\,0\,,
\end{split}
\ee}
and thus

\be
i_{\mathbb{Z}}\,\eta_{0}\,=\,\widetilde{\mathscr{F}}_{\mathbf{a}},
\ee
which means that the second equation in \eqref{eqn: equations defining the unitary+gradient vector field on the punctured hilbert space} is fulfilled.
Concerning the third equation in \eqref{eqn: equations defining the unitary+gradient vector field on the punctured hilbert space}, recalling that $\mathbb{X}_{\mathbf{a}}$ and $\mathbb{Y}^{0}_{\mathbf{b}}$ are $\pi$-related, respectively, with $X_{\mathbf{a}}$ and $Y_{\mathbf{b}}$, we have

\be
i_{\mathbb{X}_{\mathbf{a}}}\,\omega_{0}\,=\,\pi^{*}\left(i_{X_{\mathbf{a}}}\,\omega\right)\,=\,\pi^{*}\left(\mathrm{d}e_{\mathbf{a}}\right)\,=\,\mathrm{d}\widetilde{\mathscr{F}}_{\mathbf{a}}
\ee

\be
\begin{split}
i_{\mathbb{Y}^{0}_{\mathbf{b}}}\,\omega_{0}\,&=\,\pi^{*}\left(i_{Y_{\mathbf{b}}}\,\omega\right)\,=\,\pi^{*}\left(\omega\circ J(X_{\mathbf{b}})\right)\,=\\
&=\,-\pi^{*}\left(J\circ J\circ \omega \circ J(X_{\mathbf{b}})\right)\,=\\
&=\,-\pi^{*}\left(J(\mathrm{d}e_{\mathbf{b}})\right)\,=\,-\widetilde{\alpha}_{\mathbf{b}} \,,
\end{split}
\ee
and thus

\be
i_{\mathbb{Z}}\,\omega_{0}\,=\,\widetilde{\mathscr{F}}_{\mathbf{a}} - \widetilde{\alpha}_{\mathbf{b}} \,,
\ee
which means that the third equation in \eqref{eqn: equations defining the unitary+gradient vector field on the punctured hilbert space} is satisfied.
Consequently, $\mathbb{Z}$ is $\tau$-related with  a vector field $\mathcal{Z}$ on $S^{2n-1}$ which is the contact Hamiltonian vector field associated with $\mathscr{F}_{\mathbf{a}}$  and $\alpha_{\mathbf{b}}$ by means of the contact structure $(\eta_{s} \,,\omega_{s})$, and, since $\mathbb{Z}$ is $\pi$-related with $Z_{\mathbf{a},\mathbf{b}}$, it follows that $\mathcal{Z}$ is $\upsilon$-related with $Z_{\mathbf{a},\mathbf{b}}$ too.
Eventually, we obtained a description of the ``dynamical evolution'' associated with the vector field $Z_{\mathbf{a},\mathbf{b}}$ in terms of contact Hamiltonian systems as defined in Appendix A.

In this manner we have obtained evolutionary equations on the space of ``normalized'' wave functions which are not unitary and are nonlinear.
Elsewhere, we shall compare our equations with other proposals to describe dissipation on wave functions by means of nonlinear equations \cite{cruz_schuch_castanos_rosas-ortiz-time_evolution_of_quantum_systems_via_complex_nonlinear_riccati_equation_1, cruz_schuch_castanos_rosas-ortiz-time_evolution_of_quantum_systems_via_complex_nonlinear_riccati_equation_2, schuch-quantum_theory_from_a_nonlinear_perspective}.
We shall now move to the classical setting where the use of contact structures to describe dissipation has been already advanced \cite{bravetti_cruz_tapias-contact_hamiltonian_mechanics}.

\section{Preliminary considerations on second order dynamical systems}

In the following, we will exploit the idea of decomposing a dynamical system into a ``reference dynamics'' plus a ``perturbation term'' in order to give a geometrical characterization of dissipation in the classical Lagrangian setting.
The reference dynamics will be a dynamical system admitting a Lagrangian description, and the perturbation term is responsible for the ``dissipation'' of the Lagrangian energy of the reference dynamics.
Furthermore, motivated by the recent introduction of Hamiltonian contact systems (\cite{bravetti_cruz_tapias-contact_hamiltonian_mechanics}), we will consider contact dynamical systems in the context of Lagrangian mechanics and use them to describe some interesting classes of dissipative systems.  
In order to this, we start with some preliminary remarks on the description of classical Newtonian-like systems.

In the classical setting, dissipation is  usually understood as the fact that energy is not conserved along a given dynamical trajectory.
However, we want to stress that if we want to declare a system to be dissipative, we should say what is actually being dissipated. 
Indeed, it is clear that  we may deal with physical systems for which it might make sense to say that the system is ``dissipating'' mass or angular momentum or  ``probability''.
For instance, in remark \ref{remark: phase damping} we saw that the phase-damping evolution for a finite-level quantum system may be thought of as a dynamical evolution which is dissipating the off-diagonal terms of the quantum states with respect to a fixed orthonormal basis depending on the dynamical evolution itself.

When we are simply given a dynamical system whose dynamics is described by means of a vector field on a suitable carrier space, we must realize that, in addition to identifying the ``quantity'' the dissipation of which we are interested in,  it does not make sense to say that the system is ``conservative'' or ``dissipative'' {\itshape per se} because dissipation is a relational concept. 
For instance, in the Newtonian setting where dynamical vector fields are second-order vector fields on the tangent bundle $TQ$ of some configuration space $Q$, we may be interested in systems dissipating energy, and, in order to analyze such dissipation, we must first of all define and fix what we mean by ``energy of the system''.
For instance, we will see that there are dynamical systems that are usually thought of as describing some dissipative phenomenon for which it is possible to give a Lagrangian description in terms of a Lagrangian function and a Lagrangian energy function.
In this case, it is clear that the choice of some ``reference notion of energy'' carries physical information because it allows us to describe the same physical system as being conservative or dissipative.

In this section, we will deal with classical Newotnian-like systems.
These are dynamical systems for which the dynamics is expressed by means of a second-order vector field on the tangent bundle $TQ$ of some configuration-like manifold $Q$.
Roughly speaking, the fact that the vector field is of second order means that there is a coordinate system $(\mathbf{q},\,\mathbf{\dot{q}})$ on $TQ$ with respect to which the equations of motion read
\be
\frac{\mathrm{d}^{2} \mathbf{q}}{\mathrm{d}t^2}\,=\, \mathbf{F}(\mathbf{q},\,\mathbf{\dot{q}})\,.
\ee
For a more elegant and coordinate-free formulation of the notion of second-order vector field we refer to \cite{marmo_ferrario_lovecchio_morandi_rubano-the_inverse_problem_in_the_calculus_of_variations_and_the_geometry_of_the_tangent_bundle}.
Given a Newtonian dynamical system, it makes sense to ask if the ``forces'' generating the motion are derived from a potential because in this evenience we would find that the kinetic energy plus the potential one would be conserved. 
A covariant way to look for a potential is the search for a Lagrangian description.
Thus, in order to address the characterization of a given second order dynamical system as ``conservative'' or ``dissipative'', we may proceed as follows.
First of all, we look for a possible Lagrangian description of the given second order vector field.
To discuss the possibility of a Lagrange description we should formulate and solve the so-called ``inverse problem in the calculus of variations'' which has been quite extensively studied both for finite dimensional systems and fields~\cite{marmo_ferrario_lovecchio_morandi_rubano-the_inverse_problem_in_the_calculus_of_variations_and_the_geometry_of_the_tangent_bundle}, here we will not deal with this problem in complete generality, for this we refer to existing literature.
It may happen that we find none, one or many solutions for this problem, that is, either we do not find a Lagrangian description, or we find a ``unique''\footnote{As it is well-known, we may always add a global time derivative to every Lagrangian, or multiply the Lagrangian by a real number without altering the explicit form for the equation of motions, hence, uniqueness here has always to be understood modulo the addition of a global time derivative or multiplication by a real number.} Lagrangian description, or we find different alternative Lagrangian descriptions. 
Then, if a Lagrangian description exists, in the
time independent case we would say that the system preserves the Lagrangian energy function 	
	\beq\label{Lag-En}
	\mathcal{E}_{\lag} = \dot{q}_j \frac{\partial \lag}{\partial \dot{q}_j} - \lag \, ,
	\eeq
associated with the Lagrangian $\lag$ for the second order vector field $\Gamma$.
There will be an associated ``Lagrangian energy'' for each alternative Lagrangian description of the second order vector field.
However, the actual Lagrangian energy $\mathcal{E}_{\lag}$ need not coincide with what we may want to call the ``physical energy'' $E$ of the system, therefore, we could qualify the system to be dissipative even if it admits a description by means of a Lagrangian.
A simple example clarifies what we mean.
We consider a second order differential equation on $TQ={\rm I\!R}\times{\rm I\!R}$ with a friction force proportional to the velocity by means of a friction coefficient $\gamma$, i.e.

	\be
	\ddot{q} + \gamma\,\dot{q}\,=\,0\,.
	\ee
A possible (local) Lagrangian for this system with its corresponding Lagrangian energy are

	\beq
	\lag = \dot{q}(\ln{\dot{q}}) - \gamma\,q
	\quad
	\text{and}
	\quad
	\mathcal{E}_{\lag} = \dot{q} + \gamma\, q
	\eeq 
respectively. 
Then, the system allows for a Lagrangian description where the Lagrangian energy does not coincide with the mechanical energy $E = \frac{m}{2}\dot{q}^2$, and we may look at this Lagrangian dynamical system as being dissipative with respect to the mechanical energy $E$. Specifically, the rate of dissipation of the mechanical energy is $\frac{\d E}{\d t} = - \gamma\, \dot{q}^2$.

\vsp

When the system does not admit a Lagrangian description, we have to develop new strategies in order to characterize dissipation.
In order to explain such a strategy, let us consider the physically relevant situation represented by linear dynamical systems.

In several instances, either because of approximations or because of specific requirements, the equations of motion may be given to us in a linear form

	\beq \label{New-Eq}
	m^{j k} \ddot{q}_k + \gamma^{j k} \dot{q}_k + \omega^{j k}q_k = 0
	\eeq
where $||m^{j k}||$, $||\gamma^{j k}||$ and $||\omega^{j k}||$ are numerical matrices.
Notice that this equation is in general an implicit differential equation because the matrix $||m^{jk}||$ could be degenerate.
When $||m^{j k}||$ is non-degenerate, the differential equation defines a second order vector field. 
In this case, we have a necessary and sufficient condition for this vector field to admit a Hamiltonian description in terms of a constant Poisson structure on $T{\rm I\!R}^{n}$ (see chapter $4$ in  \cite{carinena_ibort_marmo_morandi-geometry_from_dynamics_classical_and_quantum}).
Specifically, consider the linear vector field $\Gamma$ associated with the equations of motion on the linear manifold $T{\rm I\!R}^{n}$, and the representative matrix $G^{j}_{k}$ of $\Gamma$ defined by

	\be
	\Gamma\,=\,G^{j}_{k}\,\xi^{k}\,\frac{\partial}{\partial \xi^{j}}\,,
	\ee
where $\{\xi^{j}\}_{j=1,...,2n}$ is a collective Cartesian coordinates system on $T{\rm I\!R}^{n}$ with the first $n$ coordinates representing position variables.
We require $G$ to be a generic matrix, i.e., with simple eigenvalues.
We have that the system may be given a description by means of a constant Poisson structure $\Lambda$ on $T{\rm I\!R}^{n}$, represented in the coordinate system $\{\xi^{j}\}_{j=1,...,2n}$ by an antisymmetric numerical matrix $||\Lambda^{jk}||$, and  a quadratic Hamiltonian function

	\beq
	H = \frac{1}{2}H_{jk}\xi^{j}\xi^{k}\, ,
	 \eeq
if and only if the representative matrix $||G^{j}_{k}||$ is traceless with all its odd powers, namely
 
	\beq
	\text{Tr}\{G^{2k + 1}\} = 0
	\quad
	\text{for all $k$}\, .
	\eeq
Even powers $G^{2k}$ will act as non-canonical symmetries of the dynamics which take from one Hamiltonian description to an alternative one. 
If the representative matrix is not generic, additional work is required to take into account its specificity, we refer to \cite{carinena_ibort_marmo_morandi-geometry_from_dynamics_classical_and_quantum} for a complete treatment.

Now, we consider the following system of two coupled differential equations representing coupled oscillations with different frequencies $\omega_k$, different damping coefficients $\gamma_k$ and different coupling constants $\kappa_{k}$ and $\delta_{k}$

	\begin{align} 
	\ddot{q}_1 + \gamma_1 \dot{q}_1 + \omega^2_1 q_1 + \kappa q_2 + \delta \dot{q}_2 & = 0 \nonumber \\
	\ddot{q}_2 + \gamma_2 \dot{q}_2 + \omega^2_2 q_2 + \kappa q_1 + \delta \dot{q}_1 & = 0 \, .
	\label{Coup-Dam-Osc}
	\end{align}
Then, the representative matrix is

	\beq\label{Rep-Mat}
	G = 
	\left(
	\begin{array}{cccc}
	 0 & 0 & 1 & 0 \\
	 0 & 0 & 0 & 1\\
	 -\omega^2_1 & -\kappa & -\gamma_1 & -\delta \\
	 -\kappa & -\omega^2_2 & -\delta & -\gamma_2
	\end{array}
	\right)\, .
	\eeq
This matrix cannot  be expressed as the product of a skew-symmetric matrix times a symmetric one because its trace is different from zero.

\vsp

We may, in full generality, prove that this system does not allow for any Lagrangian  description even if we allow the Lagrangian to be any function of position and velocity. To do that we recall a theorem on Lagrangian symplectic forms~\cite{balachandran_marmo_skagerstam_stern-supersymmetric_point_particles_and_monopoles_with_no_strings, marmo_ferrario_lovecchio_morandi_rubano-the_inverse_problem_in_the_calculus_of_variations_and_the_geometry_of_the_tangent_bundle}:

A two form $\omega$ on $TQ$ is derivable from a Lagrangian function (at least at the local level) if and only if
	\begin{enumerate}[label=\roman*)]
	
	\item $\left\langle \omega | \frac{\partial }{\partial \dot{q}_i}\wedge\frac{\partial }{\partial \dot{q}_j}
	 \right\rangle = 0 $,
	 for all $i,j$\, ,
	
	\item $\pounds_\Gamma \, \omega = 0$, for some second order vector field $\Gamma$\, ,
	
	\item $\d\, \omega = 0$.
	
	\end{enumerate}
If we take the Lie derivative of i) with respect to $\Gamma$, and taking into account ii), we obtain, after iterating $k$ times 
	\beq
	\left\langle \omega \Big| (\pounds_\Gamma)^k \left(
	\frac{\partial }{\partial \dot{q}_i}\wedge\frac{\partial }{\partial \dot{q}_j}\right) \right\rangle = 0\, .
	\eeq
Of course, from a certain $k$ onwards, one obtains equations which are $\mathcal{F}$--linear combination of the preceeding ones, so that the space spanned by the bivector fields
	\beq
	(\pounds_\Gamma)^k \left(
	\frac{\partial }{\partial \dot{q}_i}\wedge\frac{\partial }{\partial \dot{q}_j}\right) \, 
	\eeq
at each point, $(q_j,\dot{q}_j)\in TQ$ is finite dimensional, and in fact its maximum dimension is $n(2n-1)$. Then, when the dimension is maximal $\omega$ is obliged to be zero and the inverse problem has no solution.
For example, from the dynamical vector field defined by the representative matrix in Eq. \eqref{Rep-Mat} one can obtain that the set of independent bivectors is given by 
	\begin{align}
	X_0 = \frac{\partial }{\partial \dot{q}_1}\wedge\frac{\partial }{\partial \dot{q}_2}
	\qquad &\qquad 
	X_1 = \frac{\partial }{\partial q_1}\wedge\frac{\partial }{\partial q_2}  \nonumber\\
	X_2 = \frac{\partial }{\partial q_1}\wedge\frac{\partial }{\partial \dot{q}_1}
	\qquad &\qquad 
	X_3 = \frac{\partial }{\partial q_1}\wedge\frac{\partial }{\partial \dot{q}_2} \nonumber\\
	X_4 = \frac{\partial }{\partial q_2}\wedge\frac{\partial }{\partial \dot{q}_1}
	\qquad &\qquad 
	X_5 = \frac{\partial }{\partial q_2}\wedge\frac{\partial }{\partial \dot{q}_2} \, .
	\end{align}
Therefore, the dimension is maximal and the Lagrangian does not exist.

If the system does not admit a Lagrangian description, we may proceed in analogy with the quantum case described in section \ref{section: Geometrical description of the GKLS equation and dissipation} and manipulate the system of differential equations in \eqref{Coup-Dam-Osc} in order to obtain a particularly useful decomposition of the dynamical system it describes.
For instance, it is easy to see that the representative matrix of the system given in Eq.~\eqref{Rep-Mat} can be decomposed as $||G^{j}_{k}|| = ||A^{j}_{k}|| + ||D^{j}_{k}||$, where $||A^{j}_{k}||$ is a traceless matrix, while $||D^{j}_{k}||$ is a ``remainder''.
Clearly, the decomposition into a traceless matrix  and the ``remainder'' is arbitrary.
Once a choice has been made, it is possible to think of the dynamical vector field as the sum of a ``comparison'' or ``reference dynamics'' with a ``perturbation term''.	 
We may select the ``comparison dynamics'' in such a way that it admits at least one Lagrangian description.
Then, the perturbation term will turn our system into a ``dissipative'' system which is dissipating the Lagrangian energy of the comparison system.
Eventually, the linear vector field $\Gamma$ may be decomposed as

	\be
	\Gamma\,=\,A^{j}_{k}\,\xi^{k}\,\frac{\partial}{\partial \xi^{j}} + D^{j}_{k}\,\xi^{k}\,\frac{\partial}{\partial \xi^{j}}\,,
	\ee
where the first term on the right-hand side plays the role of the ``comparison dynamics'' admitting a Hamiltonian description, while the second term may be thought of as the perturbation term responsible for the dissipation of the Hamiltonian energy function associated with the comparison dynamics. 
This decomposition is analogous to the one we have described for the GKLS vector field.
This analysis can be extended to the general linear systems in Eq.~\eqref{New-Eq}, however, it is important remark that the decomposition is not going to be unique and thus we will have many alternative ``energy functions'' which are going to be dissipated by the same perturbation.

Once a comparison system has been selected, it is possible to go to a contact formalism to show that a ``dissipative'' system may be described in terms of a ``contact Hamiltonian formalism''.
One may hope that, having a kind of Hamiltonian formalism along with a possible Hamilton--Jacobi picture, the way would be paved to arrive also at writing a ``dissipative'' Schr\"{o}dinger-type equation not so much as a genuine ``quantization'' but rather as quantum equation which in a suitable ``classical'' limit would reproduce the classical ``dissipative system'' we had started from.

\section{Contact Lagrangian structures and dissipations}
\label{section: Contact Lagrangian structures and dissipations}

Once a ``decomposition principle'' for the system has been defined, it is possible to go to a contact formalism to show that a ``dissipative'' system may be described in terms of a ``contact Hamiltonian formalism''.
The framework we will refer to is the formalism of contact Hamiltonian systems developed in Appendix A which is an extension of the formalism presented in \cite{bravetti_cruz_tapias-contact_hamiltonian_mechanics} to the case of contact manifolds that need not be exact.

In particular, assuming that the carrier space for the dynamics is  $TQ \times {\rm I\!R}$ with an exact contact structure $(\eta, \, \xi)$, in order to define the dynamics on  $TQ \times {\rm I\!R}$ one can associate with every smooth function $\E$ a vector field $\Gamma_{\tiny \mbox{C}}$ on $TQ \times {\rm I\!R}$  by means of
	\beq\label{Diss-Lag-G}
	i_{\Gamma_{\tiny \mbox{C}}} \d \eta = \d \E - (\pounds_\xi \E) \eta
	\qquad
	\text{and}
	\qquad
	i_{\Gamma_{\tiny \mbox{C}}} \eta = - \E \, ,
	\eeq
where $\E$ is called the ``contact Lagrangian energy". 

In particular, we assume that  the $1$-form $\eta$ can be written as
	\beq \label{one-form}
	\eta = \d S  -  \theta_\lag 
	\quad
	\text{with}
	\quad
	 \theta_\lag =  \d q_j \frac{\partial\lag}{\partial \dot{q}_j} 
	  \, ,
	\eeq
where $(q_{j},\,\dot{q}_{j},\,S)$ are local coordinates on $TQ \times {\rm I\!R}$, $\lag$ is the Lagrangian function of the ``comparison system", and the contact Lagrangian energy $\E\equiv \E_{\lag}$ may be written as
	\beq \label{Cont-En}
	\E_\lag = \mathcal{E}_{\lag} + h(S)\, .
	\eeq
The first term $\mathcal{E}_\lag$ is the Lagrangian energy in Eq.~\eqref{Lag-En} of the conservative ``comparison system'', while the second term $h(S)$ is a ``perturbation" on the system giving an effective characterization of the interaction between the conservative system and the environment. 
The system is thus decomposed into a ``conservative (Lagrangian) comparison dynamics'' and a ``dissipative term'', and what is being dissipated is the Lagrangian energy of the conservative comparison dynamics. 

Introducing definitions \eqref{one-form} and \eqref{Cont-En} into the conditions in \eqref{Diss-Lag-G} we obtain the Euler--Lagrange equations and the equation for the component of the vector field in the direction of the Reeb vector, namely
	\begin{align}
	\pounds_{\Gamma_{\tiny \mbox{C}}} \theta_{\lag} - \d \lag & = - \frac{\d h}{\d S} \,\theta_\lag\, , \\
	\dot{S} & = i_{\Gamma_{\tiny \mbox{C}}} \theta_\lag - \E_\lag \, .
	\end{align}
Equivalently, using the definition for the $1$-form $\theta_\lag$ in Eq. \eqref{one-form} and spelling the Lie derivative on $\theta_\lag$ it is easy to obtain the coordinate expression of the Euler--Lagrange equations
	\beq \label{Cont-EL-Eq}
	\frac{\d}{\d t} \frac{\partial \lag}{\partial \dot{q}_j} 
	- \frac{\partial \lag}{\partial q_j} 
	= - \frac{\d h}{\d S} \frac{\partial \lag}{\partial \dot{q}_j}\, ,
	\eeq	
which in general are implicit differential equations. In addition, we have in coordinates that
	\beq
	\dot{S} =\lag - h(S)\, .
	\eeq
	
We may look at these systems as a sort of generalization of the so-called Caldirola-Kanai dissipative systems.
Here, the Lagrangian energy is not preserved along the dynamical trajectories, indeed
 	\beq
	\frac{\d \mathcal{E}_{\lag}}{\d t} = - \frac{\d h}{\d S} \frac{\partial \lag}{\partial \dot{q}_j}\, ,
	\eeq
which may be positive or negative according to the sign of $\frac{\d h}{\d S}$. Additionally, the vector field associated with the dynamical system can be easily decomposed into a  comparison dynamics admitting a Lagrangian description and a perturbation term (contact corrections). 

Defined the contact dynamics $\Gamma_{\tiny \mbox{C}}\in\X(TQ \times {\rm I\!R})$ a natural question is whether it is possible to project such dynamics onto a second order vector field $\Gamma \in \X(TQ)$ to relate such dynamics  with a second order dynamics in the original variables. 
This means that we get an ``effective'' equation of motion on the original variables $(\mathbf{q},\,\dot{\mathbf{q}})$ which is ``decoupled'' from the equation on the ausiliary variable $S$.
Then, it is clear from contact Euler--Lagrange Eq. \eqref{Cont-EL-Eq}, that it reproduces a second order dynamics if $h(S)$ is proportional to $S$. 
Formally, a contact dynamics $\Gamma_{\tiny \mbox{C}}$ is projectable onto a second order dynamics if and only if $[\Gamma_{\tiny \mbox{C}}, \xi]\wedge \xi = 0$, in addition, 
this directly implies that the function $\E\in \F(TQ \times {\rm I\!R})$, associated to the contact vector field, is linear in $S$. 
When the contact Lagrangian energy is not linear in the $S$ variable we have a dissipative system that although not projectable, may still describe physical systems see for example~\cite{grmela_ottinger-dynamics_and_thermodynamics_of_complex_fluids_I}.

\begin{figure}[h!]
\begin{center}
\includegraphics[scale=0.2]{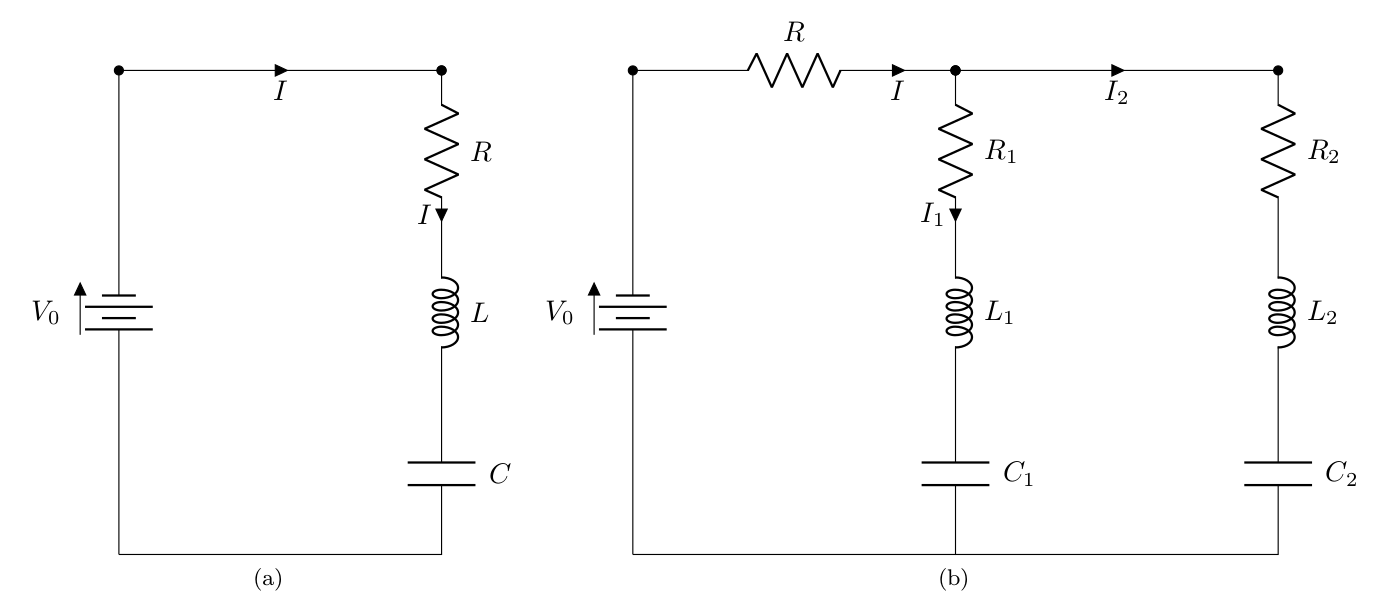}
\caption{(a) Diagram of a basic RLC circuit. (b) Diagram of the coupling of two RLC circuits connected 		in parallel by means of the resistance R. }
		\label{RCL-1}
\end{center}
\end{figure}
%

Concrete examples of second order linear systems which have a clear physical interpretation are provided by electric circuits. For instance, for the description of the basic circuit in Fig.~\ref{RCL-1}a we can consider 
	\beq
	\lag = \frac{1}{2}L \, \dot{I}^2 - \frac{1}{2 \, C} I^2 
	\quad
	\text{and}
	\quad
	h(S) = R \, S\, , 
	\eeq
using the notation indicated in the figure. Thus, the Euler--Lagrange equation \eqref{Cont-EL-Eq} for this case corresponds to the usual RLC equation
	\beq
	L\, \ddot{I} + R \dot{I} + \frac{1}{C} I = 0\, .
	\eeq

It is clear that this generalization of the Caldirola-Kanai dissipation allows to describe dissipative systems which are ``linear in the velocity''. 
However, in order to handle more general situations like the one presented in Fig.~\ref{RCL-1}(b), it is necessary to extend the formalism and consider a more general instance of contact structure. As is exposed in Appendix A, it is possible to define in the more general sense a contact manifold by means of a global
1-form $\eta$ and a global 2-form $\omega$. In particular, let us take our contact manifold to be again  $TQ \times {\rm I\!R}$. 
Then, we define a dynamical evolution in terms of a vector field associated with a smooth function $\mathscr{\E_\lag}$ by means of
	\beq\label{eqn: contact E-L}
	i_{\Gamma_{\tiny \mbox{C}}} \omega = \d \E - (\pounds_\xi \E) \eta
	\qquad
	\text{and}
	\qquad
	i_{\Gamma_{\tiny \mbox{C}}} \eta = - \E \, .
	\eeq
Now, similarly to what we have done in the case of exact contact structures on $TQ \times {\rm I\!R}$, here we assume that $\eta$ and $\omega$ are given by

	\beq
	\eta = \d S  - \alpha 
	\qquad
	\text{and}
	\qquad
	 \omega = - \, \d \theta_\lag 
	  \, ,
	\eeq
where $\alpha$ is a semi-basic $1$-form with the following local expression

	\beq
	\alpha = \sum_{i=1}^{n} a_j(q_j,\dot{q}_j,S) \, \d q_j \, .
	\eeq 
Because we are interested in characterizing dissipative systems in terms of a ``decomposition principle'', we shall assume that the Lagrangian energy $\E_\lag$ has the form defined in \eqref{Cont-En}. 
Inserting these definitions into equation \eqref{Cont-vect-field} we get the conditions

	\beq\label{general-cond}
	\pounds_{\Gamma_{\tiny \mbox{C}}} \theta_{\lag} - \d \lag = - \frac{\d h}{\d S} \,\alpha
	\qquad
	\text{and}
	\qquad
	\dot{S} = i_{\Gamma_{\tiny \mbox{C}}}\alpha - \E_\lag\, .
	\eeq
The first condition in \eqref{general-cond} determines uniquely $\Gamma_{\tiny \mbox{C}}$ up to a vector field proportional to $\xi$, and this additional term is fixed by the second requirement in \eqref{eqn: contact E-L}. 
Because we want $\Gamma_{\tiny \mbox{C}}$ to be projectable onto a second order dynamics $\Gamma \in \X(TQ)$, then it is necessary and sufficient that $h(S)$ be linear in $S$ and $\alpha$ must be a semi-basic $1$-form independent of $S$.
Therefore, we can associate the contact dynamics with a second order differential equations with respect to the $q$ variables in $TQ$.

We may identify some special classes of dissipative systems by making qualifications for $\alpha$.
For instance, we may take

	\beq
	\alpha = \d_\S \mathcal{F} \, .
	\eeq
where $\FF$ is an arbitrary velocity-dependent function, and the operator $\d_S$ is the exterior derivative associated with the (1, 1)-type tensor given by $\S = \frac{\partial }{\partial \dot{q}_j}\otimes \d q_j$

\be
(\d_\S f)(\Gamma)\coloneqq \d f(\S[\Gamma])\,.
\ee
This situation gives rise to a conformal version of the Rayleigh dissipation where the rate of change of the Lagrangian energy is

	\beq
	\frac{\d \mathcal{E}_{\lag}}{\d t} = - \frac{\d h}{\d S} \frac{\partial \FF}{\partial \dot{q}_j}\, .
	\eeq

We can now employ the conformal version of the Rayleigh dissipation for the description of the circuit in Fig.~\ref{RCL-1}b. This consists in the composition of two RLC circuits connected in parallel, where the coupling is carried out by a resistance $R$. Then, the dynamics of the system is determined by
	
	\beq
	\LL = \frac{1}{2}  L^{jk} \dot{I}_k \dot{I}_j - \frac{1}{2} C^{jk} I_k I_j \, ,
	\quad
	h(S) = S
	\quad
	\text{and}
	\quad
	\FF = - \frac{1}{2} R^{jk} \dot{I}_k \dot{I}_j\, ,
	\eeq
where here $I_1$ and $I_2$ denotes the currents in the branches and according to the notation in Fig.~\ref{RCL-1}b the matrices are
	\beq
	L =
	\left(
	\begin{array}{cc}
	 L_1 & 0  \\
	 0 & L_2
	\end{array}
	\right)\, ,
	\quad
	C =
	\left(
	\begin{array}{cc}
	 1/C_1 & 0  \\
	 0 & 1/C_2
	\end{array}
	\right)
	\quad
	\text{and}
	\quad
	R =
	\left(
	\begin{array}{cc}
	 R_1 & R  \\
	 R & R_2
	\end{array}
	\right) \, .
	\eeq
Therefore, the Euler--Lagrange equations associated with this system corresponds to
	\beq
	L^{jk} \ddot{I}_k + R^{jk} \dot{I}_k + C^{jk} I_k = 0\, ,
	\eeq
which are in agreement with the equations obtained from the Kirchhoff's circuit laws. In addition, it is clear that if the coupling parameter $R \to 0$ the system is reduced to two non-interacting RLC circuits.

\section{Conclusions}

Motivated by the description of ``dissipation'' for quantum systems in terms of GKLS equations we have argued that  the analogue of this equation on the space of pure quantum states is associated with a nonlinear evolution equation on the Hilbert space.
If we impose that the evolution should preserve the normalization of the ``wave function'', we end up with an evolution on the contact manifold of normalized vectors.
On the contact manifold, the Poisson brackets are no more available, the analogue of Hamiltonian vector fields will be the ``contact Hamiltonian vector fields''.
With the help of the complex structure on the Hilbert space we have been able to define vector fields which contain a ``conservative'' part and a ``dissipating'' part. 
While we postpone a comparison analysis with existing proposals for dissipative Schr\"{o}dinger equations, we have considered the corresponding classical situation where it is easier to compare with existing proposals because in both cases we are dealing with finite dimensional manifolds.
In particular, we have found it necessary to deal with contact structures in the Lagrangian setting.
Here we have found the necessity to go beyond the proposed description in terms of contact Hamiltonian vector fields as in \cite{bravetti_cruz_tapias-contact_hamiltonian_mechanics} already for a simple RLC circuit.
We plan to deal with GKSL dynamics on mixed states with the help of their representation on phase space by means of Wigner functions.
On phase space we may consider corresponding classical systems within the Koopman approach \cite{koopman-hamiltonian_systems_and_transformations_in_hilbert_space}, but all of this should appear in the near future.



\section*{Appendix A. Contact Hamiltonian systems}

\label{appendix-A}

Here, we will provide an extension of the notion of contact Hamiltonian system given in \cite{bravetti_cruz_tapias-contact_hamiltonian_mechanics} to the general framework of contact manifolds that need not be exact.
The contact geometry is the odd-dimensional counterpart of the symplectic geometry. 
A $(2n+1)$-dimensional manifold $M$ is said to be an \emph{exact contact manifold} or to carry a contact structure if it carries a global differentiable $1$-form $\eta$  such that
	\begin{equation}\label{int-exc-cont}
	\eta \wedge (\d \eta)^n \neq 0\,,
	\end{equation}
everywhere on $M$, we call $\eta$ a \emph{contact form}. 
In addition, the left hand side in \eqref{int-exc-cont} provides the \emph{standard volume form} on $M$, i.e. the contact manifold in this sense is orientable. 

From a more geometrical point of view, a contact structure on a manifold may be thought of in terms of a \emph{hyperplane field} on $M$, i.e. by means of a $2n$-dimensional sub-bundle $\mathscr{D}$ of the tangent bundle $TM$. 
The hyperplane is often called a \emph{Pfaffian equation}. 
A Pfaffian equation on $M$, is a codimension one distribution $\mathscr{D}: M \to TM$. 
The subspace $\mathscr{D}^\perp \subset T^\ast M$, defined by
	\beq
	\mathscr{D}^\perp = \{ \eta \in T^\ast M  \}
	\eeq
is one dimensional and locally it can be represented by a $1$-form $\eta$. 
The class of $\mathscr{D}$ at $m$ is the integer $C(m) = \max \{2p+1| \alpha \wedge (\d \alpha)^p \neq 0 \, \, \mbox{at} \, \, m  \}$. A point $m\in M$ is said to be a singular point of $\mathscr{D}$ if $C(m) < n $. Therefore a contact structure in the wider sense is a codimension one distribution without singularities. 

The condition \eqref{int-exc-cont} implies that $\mathscr{D}$ is not integrable, actually it is as far as possible from being integrable. $\mathscr{D}$ is also called \emph{the contact distribution}. 
Moreover, the orientability of $M$ and the regularity of $\mathscr{D}$ implies that the line bundle $TM/\mathscr{D}$ admits a cross section $s$ on which $\langle \eta | s \rangle = 1$. 
Thus $M$ admits a global non-vanishing vector field, say $\xi$, such that
	\beq
	i_{\xi} \eta = 1, \quad i_{\xi} \d \eta = 0 \, ,
	\eeq
$\xi$ is called the characteristic vector field of the contact structure (or also the Reeb vector field). Two basic properties of $\xi$ are the invariance of $\eta$ and $\d \eta$ under its $1$-parameter group, i.e.
	\beq
	\pounds_{\xi} \eta = 0
	\quad 
	\rm{and}
	\quad
	\pounds_{\xi} \d \eta = 0 \, .
	\eeq
Finally, we say that a contact structure is regular if $\xi$ defines a regular distribution, i.e. 
$M\xrightarrow{\pi_\xi} M/ \{ \xi \}$ is a smooth projection onto the quotient $M/ \{ \xi \}$.

From the definition of the contact structure by means of a contact distribution we can define a contact manifold in a more general sense. 
In general, a contact manifold is defined as an odd-dimensional differential manifold admitting a global $1$-form $\eta$ and a global $2$-form $\omega$ such that 
	\beq\label{non-int}
	\eta \wedge \omega^n \neq 0
	\eeq
everywhere.
Similarly to what happens in the case of exact contact manifolds, the $(2n+1)$-form $\Omega=\eta \wedge \omega^n$ provides a volume form. It is clear that the exact contact manifold is a special case with $\omega=\d \eta$.

\begin{remark}[Contact manifolds and Jacobi brackets]

Given a manifold with contact structure $(\eta, \omega, \xi)$ it is possible define a Lie algebra structure of the space of functions by means of \cite{asorey_ciaglia_dicosmo_ibort_marmo-covariant_jacobi_brackets_for_test_particles}
	\beq
	[\mathscr{F}, \mathscr{G}] \, \Omega = (n - 1) \, (\d \mathscr{F} \wedge \d \mathscr{G} \wedge \eta) \wedge
	\omega^{n-1} 
	+ (\mathscr{F} \, \d \mathscr{G} - \mathscr{G} \, \d \mathscr{F} )\, \omega^n   
	\eeq
which is called the Jacobi brackets. 
Although it defines a Lie algebra, i.e., antisymmetry and Jacobi identity are satisfied, the Leibnitz rule is not satisfied in general. In a more standard form the Jacobi brackest  can be expressed as
	\beq
	[\mathscr{F}, \mathscr{G}] = \mathscr{F} (\pounds_\xi \, \mathscr{G}) 
	+ \mathscr{G} (\pounds_\xi \, \mathscr{G}) 
	+ \Lambda(\d \mathscr{F}, \d \mathscr{G})
	\eeq
where $\Lambda$ is the bivector associated to the two-form $\omega$ and $\xi$ preserves $\Lambda$. 
Therefore, in general, one may define the contact Hamiltonian vector field $\Gamma$ associated to the Hamiltonian function $\mathscr{F}$ by means of
	\beq
	\Gamma
	=  \mathscr{F} \xi
	+ \Lambda(\d \mathscr{F}, \, \cdot \,)\, ,
	\eeq
and this association is a homomorphism of Lie algebra, i.e.
	\beq
	[\Gamma_\mathscr{F}, \Gamma_\mathscr{G}] 
	= \Gamma_{[\mathscr{F},\mathscr{G}]}\, .
	\eeq
\end{remark}

At this point, inspired by the symplectic formulation of Hamiltonian mechanics, we may define a dynamical vector field on a contact manifold as the vector field $\Gamma$ which is associated with the smooth function $\mathscr{F}$ by means of
	\beq \label{Cont-vect-field}
	i_{\Gamma} \, \Omega = n \, (\d \mathscr{F} \wedge \eta) \wedge \omega^{n-1} 
	+ \mathscr{F} \, \omega^n \, .
	\eeq
We call any such dynamical system a contact Hamiltonian system.
It is not difficult to see that for $\mathscr{F} = 1$ we obtain that $\Gamma_{\tiny \mbox{C}}$ corresponds to the Reeb vector field. it is not difficult to realize that definition \eqref{Cont-vect-field} is equivalent to
	\beq \label{Cont-vect-field 2}
	i_\Gamma \,\omega = (\pounds_\xi \mathscr{F}) \eta - \d \mathscr{F} 
	\qquad
	\text{and}
	\qquad
	i_\Gamma \eta = \mathscr{F} \, .
	\eeq
In section \ref{section: Contact Lagrangian structures and dissipations} we have shown how to write well-known classical dissipative systems in terms of  contact Hamiltonian vector fields according to equations  \eqref{Cont-vect-field} and \eqref{Cont-vect-field}.

We may take a step further and define a generalized contact Hamiltonian vector field on a contact manifold as the unique vector field $\Gamma$ associated with the smooth function $\mathscr{F}$ and the one-form $\alpha$ by means of
	\beq \label{Cont-vect-field 3}
	i_{\Gamma} \, \Omega = n \, ((\d \mathscr{F} -\alpha)\wedge \eta) \wedge \omega^{n-1} 
	+ \mathscr{F} \, \omega^n \, .
	\eeq
We call any such dynamical system a generalized contact Hamiltonian system.
It is not difficult to see that equation \eqref{Cont-vect-field 3} is equivalent to

	\beq \label{Cont-vect-field 4}
	i_\Gamma \,\omega = (i_\xi \mathscr{F} - i_{\xi}\alpha) \eta - \d \mathscr{F} + \alpha 
	\qquad
	\text{and}
	\qquad
	i_\Gamma \eta = \mathscr{F} \, .
	\eeq
In section \ref{section: Contact structures and dissipative-like systems for pure quantum states} we show how generalized contact Hamiltonian vector fields appears in the context of quantum mechanical-like systems.

\section*{Acknowledgements}

G.M. acknowledge financial support from the Spanish Ministry of Economy and Competitiveness, through the Severo Ochoa Programme for Centres of Excellence in RD (SEV-2015/0554). 
G.M. would like to thank for the support provided by the Santander/UC3M Excellence Chair Programme 2016/2017.
H.C. is grateful for the scholarship provided by CONACyT M\'exico, with reference number 379177.

\end{document}